\documentclass[final,numberedheadings]{aipproc}
\layoutstyle{6x9}
\newcommand{\be}{\begin{equation}}
\newcommand{\ee}{\end{equation}}
\newcommand{\ba}{\begin{eqnarray}}
\newcommand{\ea}{\end{eqnarray}}
\newcommand{\la}{\langle}
\newcommand{\ra}{\rangle}
\newcommand{\di}{ {\rm d} }

\begin{document}

\title{\hspace{-0.1cm}Sivers effect at HERMES, COMPASS \& CLAS12\hspace{-0.2cm}}


\author{S.~Arnold}{
  address={Institut f\"ur Theoretische Physik II, Ruhr-Universit\"at Bochum, Germany}}

\author{A. V. Efremov}{
  address={Joint Institute for Nuclear Research, Dubna, 141980 Russia}}

\author{K. Goeke}{
  address={Institut f\"ur Theoretische Physik II, Ruhr-Universit\"at Bochum, Germany}}

\author{M. Schlegel}{
  address={Theory Center, Jefferson Lab, Newport News, VA 23606, USA}}

\author{P. Schweitzer}{
  address={Institut f\"ur Theoretische Physik II, Ruhr-Universit\"at Bochum, Germany}}

\begin{abstract}
Single spin asymmetries in semi-inclusive deep-inelastic scattering off transversely
polarized targets give information on, among other fascinating effects, a pseudo
time-reversal~odd parton distribution function, the 'Sivers function'. In this
proceeding
\footnote{Talk given at the CLAS 12 RICH Detector Workshop, January 28 - 29, 2008, Jefferson Lab.}
 we review the extractions of this function from HERMES and COMPASS data.
In particular, the HERMES pion and kaon data suggest significant sea-quarks contributions 
at $x\simeq0.15$ to the Sivers effect. We present a new fit that includes
all relevant sea quark distributions and gives a statistically satisfactory overall
description of the data, but does not describe ideally the $K^+$ data from HERMES.
We argue that measurements of the pion- and kaon Sivers effect at CLAS12,
and COMPASS, will clarify the situation.
\end{abstract}

\maketitle


\section{Introduction}

Ever since the first large single spin asymmetries (SSA)
were observed in hadron-hadron collisions
\cite{Bunce:1976yb,Apokin:1988sn,Adams:1991rw,Adams:1991cs}
spin phenomena in QCD became more and more popular.
In particular the SSA in semi-inclusive deeply inelastic scattering (SIDIS)
from transversely polarized targets have recently been measured at HERMES and
COMPASS \cite{Airapetian:2004tw,Alexakhin:2005iw,Diefenthaler:2005gx,Ageev:2006da,Diefenthaler:2006vn,Diefenthaler:2007rj,Martin:2007au,Vossen:2007mh,Alekseev:2008dn,Kotzinian:2007uv}.
SSA in SIDIS with longitudinally polarized leptons or nucleons have been reported
at HERMES and the Jefferson Lab
\cite{Airapetian:1999tv,Airapetian:2001eg,Airapetian:2002mf,Airapetian:2005jc,Avakian:2003pk,Avakian:2005ps,Airapetian:2006rx}.
In a partonic picture structure functions in SIDIS and the Drell-Yan (DY)
process with small transverse momenta $P_{\perp}\ll Q$ of the final state
hadron or the lepton pair, respectively, are sensitive to the intrinsic
transverse motion of partons \cite{Ralston:1979ys,Mulders:1995dh,Boer:1997nt},
see also the review \cite{Bacchetta:2006tn}. This means in the case of SIDIS
that structure functions can be expressed in terms of convolutions of transverse
momentum dependent (TMD) parton distributions and fragmentation functions.
Strict factorization formulae for these processes involving an additional soft
factor were discussed and established in \cite{Ji:2004wu,Ji:2004xq,Collins:2004nx}.

In the production of unpolarized hadrons in DIS of an unpolarized lepton beam off a
transversely polarized target there are three leading twist spin structure functions,
see e.g.\ \cite{Bacchetta:2006tn}, which can be distinguished by their azimuthal
distributions proportional to $\sin(\phi+\phi_s)$, $\sin(\phi-\phi_s)$, and
$\sin(3\phi-\phi_s)$. Here $\phi$ denotes the angle between lepton plane and
hadron plane in the lab frame while $\phi_s$ represents the angle of the transverse
spin vector with respect to the lepton plane. In the parton model the
structure function proportional to  $\sin(\phi+\phi_s)$ (``Collins effect'')
is expressed in terms of two chirally-odd correlation functions, namely the
transversity parton distribution $h_1^a$ which is not accessible in inclusive DIS,
and the Collins fragmentation function $H_1^\perp$ \cite{Collins:1992kk}.
The structure function proportional to $\sin(\phi-\phi_s)$ (``Sivers effect'')
is described by the T-odd parton distribution $f_{1T}^{\perp}$, the so-called
Sivers function \cite{Sivers:1989cc,Sivers:1990fh}, in conjunction with the usual
unpolarized fragmentation function $D_1$. The structure function proportional to
$\sin(3 \phi-\phi_s)$ provides information on the so-called "pretzelosity"
distribution $h_{1T}^{\perp}$ \cite{Mulders:1995dh} whose physical interpretation
was recently discussed in \cite{Miller:2007ae}.

\begin{figure}[t!]
\includegraphics[width=8cm]{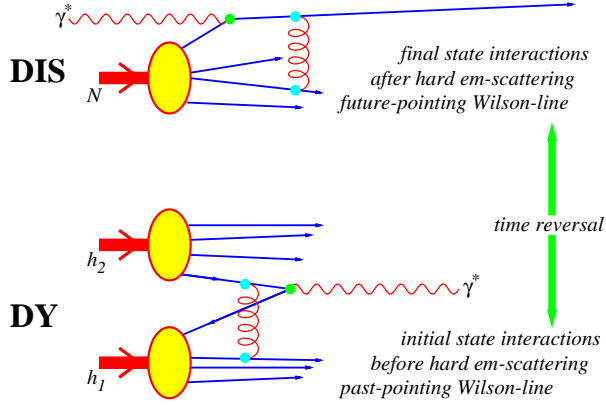}
\caption{\label{fig1-Sivers-in-DIS-vs-DY}
    The SSA due to Sivers effect arises in SIDIS from
    final state interactions \cite{Brodsky:2002cx} (upper part), and in DY from
    initial state interactions \cite{Brodsky:2002rv} (lower part). Both types
    of interactions are encoded appropriately defined Wilson lines that are
    connected to each other by time reversal \cite{Collins:2002kn}.
    In the Figure the respective interactions are sketched in the
    one-gluon-exchange approximation, see text.}
\end{figure}

T-odd parton distributions such as the Sivers and also Boer-Mulders function
$h_1^{\perp}$ \cite{Boer:1997nt} were considered to vanish due to time-reversal
symmetry for some time.
Only when it became clear that initial/final state interactions between the
struck quark and the target remnants in the parton model can cause SSAs
\cite{Brodsky:2002cx,Brodsky:2002rv}, see Fig.~\ref{fig1-Sivers-in-DIS-vs-DY},
the existing definitions of transverse momentum dependent parton distributions
were revisited. It was then shown that initial/final state interactions can be
implemented into the definitions of parton distributions by means of gauge-link
operators with appropriate Wilson lines
\cite{Collins:2002kn,Belitsky:2002sm,Ji:2002aa,Boer:2003cm}. In particular
these Wilson lines ensure the color gauge invariance of the definition of TMD
parton distributions and fragmentation functions. The application of
time-reversal switches the direction of the Wilson lines from future-pointing
(final state interactions) to past-pointing (initial state interactions) lines
in conjunction with an overall sign change. Therefore T-odd parton distributions
do not vanish. Instead, time-reversal establishes a connection between T-odd parton
distributions in processes with final state interaction (e.g. SIDIS) and  in
processes with initial state interactions (e.g. Drell-Yan). For the Sivers
function the relation reads
\begin{equation}\label{eq:Siversrel}
     f_{1T}^{\perp}(x,\vec{k}_T^2)\Big|_{\mathrm{SIDIS}}\, = \,
    -f_{1T}^{\perp}(x,\vec{k}_T^2)\Big|_{\mathrm{DY}}.
\end{equation}
This important QCD-prediction can and still needs to be checked by experiments.

In these proceedings we discuss the present status of the understanding of the
Sivers function from HERMES and COMPASS \cite{Efremov:2004tp,Anselmino:2005ea,Vogelsang:2005cs,Collins:2005ie,Collins:2005rq,Collins:2005wb,Anselmino:2005an,Efremov:2007kj,Efremov:2008vf}.
In particular we review the works
\cite{Efremov:2004tp,Collins:2005ie,Collins:2005rq,Collins:2005wb,Efremov:2007kj,Efremov:2008vf}
and discuss the developments due to the most recent data from SIDIS.
We also present predictions for pion and kaon Sivers asymmetries for the 12 GeV
upgrade at Jefferson Lab.

\section{Semi-inclusive processes}

\begin{figure}[b!]
\begin{tabular}{cc}
\includegraphics[height=4cm]{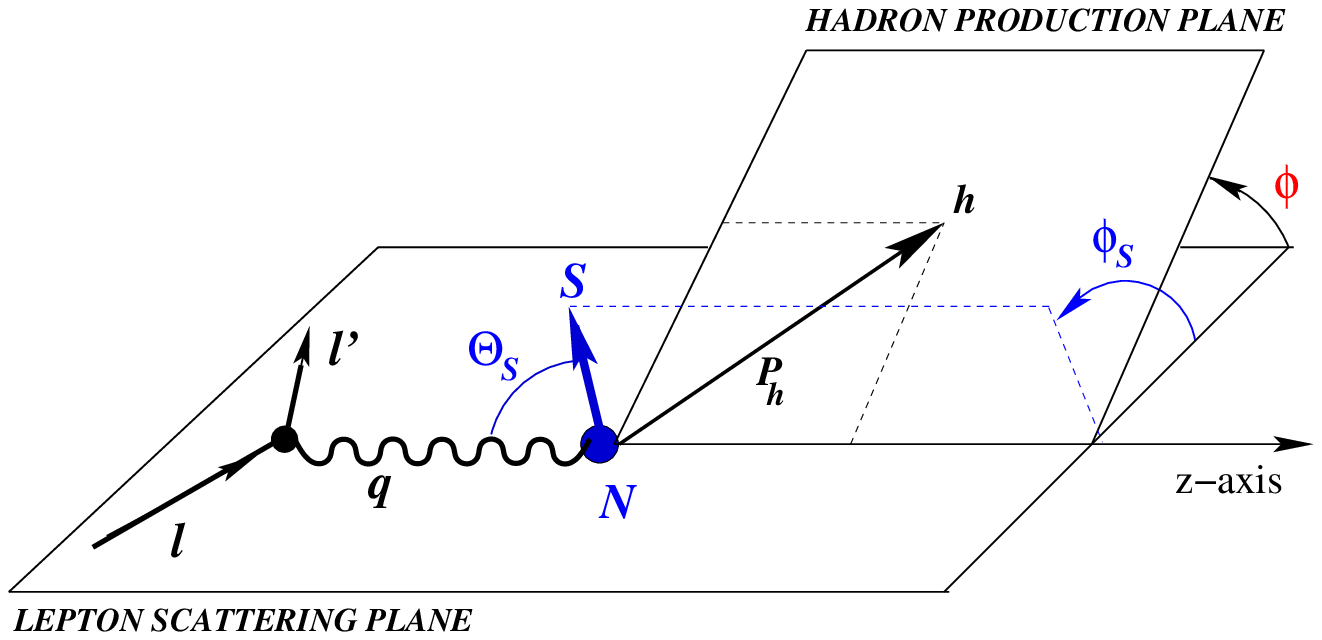} &
\includegraphics[height=4cm]{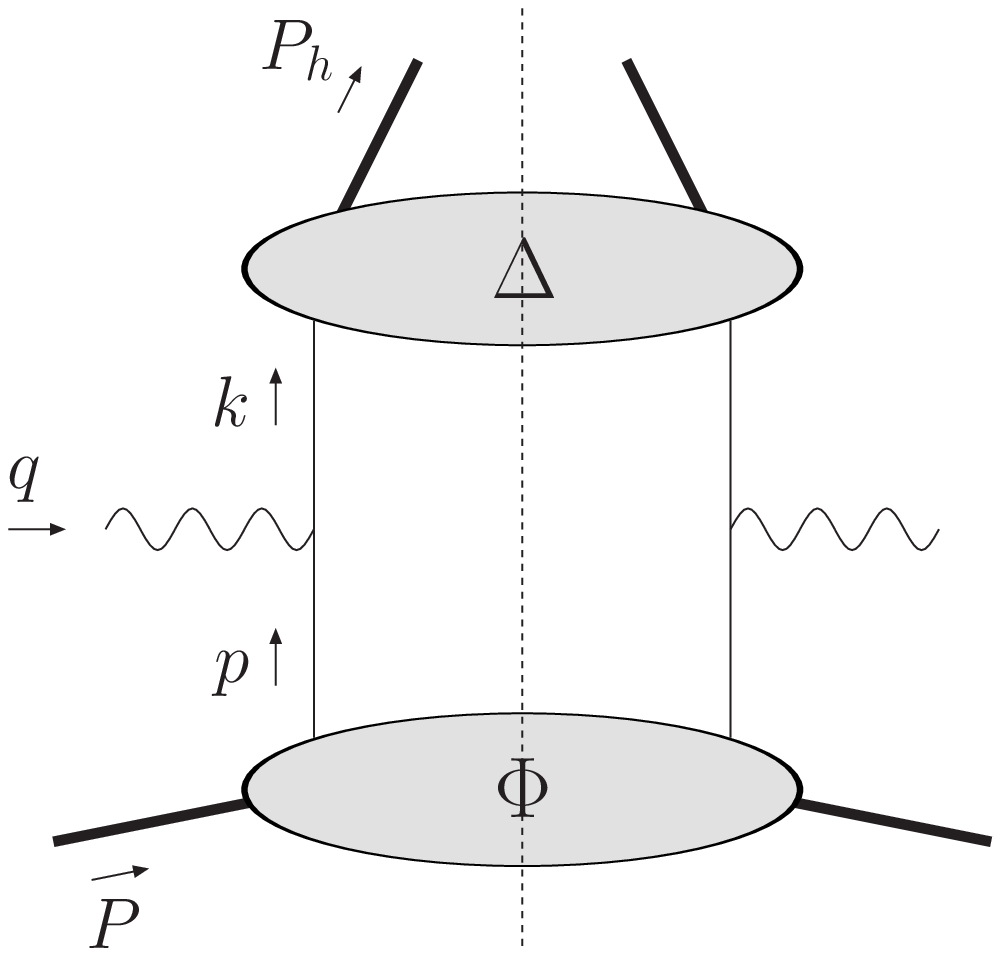}
\end{tabular}
    \caption{
    \label{fig2-kinematics}
    Left panel:
    The kinematics of the SIDIS process $lN\rightarrow l^\prime h X$.
    The nucleon is polarized transversely with respect to the beam.
    However, up to power corrections the polarization is transverse
    also with respect to the momentum of the virtual photon:
    $\sin\Theta_S\sim M_N/Q\ll 1$.
    Right panel:
    Tree-level diagram for the hadronic tensor in leading order in
    $1/Q$ in the parton model.}
\end{figure}

In the following we briefly discuss SIDIS with transversely polarized targets,
with particular emphasis on the Sivers effect. In SIDIS of an unpolarized lepton
beam off a transversely polarized nucleon target one can measure the SSAs
\cite{Gourdin:1973qx,Kotzinian:1994dv,Diehl:2005pc,Bacchetta:2006tn}
\begin{eqnarray}
    \frac{\;\;\,  d\sigma^\uparrow-d\sigma^\downarrow }
         {\frac12(d\sigma^\uparrow+d\sigma^\downarrow)}  =
    S_T \Big[\sin(\phi_h-\phi_s)\, A_{UT}^{\sin(\phi_h-\phi_s)}
    + \varepsilon \sin(\phi_h+\phi_s)\, A_{UT}^{\sin(\phi_h+\phi_s)} + \dots\Big]
    \;.\label{eq:structuref}
\end{eqnarray}
The SSAs $A_{UT}^{w(\phi,\phi_S)}\equiv F_{UT}^{w(\phi,\phi_S)}/F_{UU}$ are
ratios of the respective structure functions to the unpolarized one, which
depend on $x=Q^2/(2P\cdot q)$, $z=(P\cdot P_h)/(P\cdot q)$, $\vec{P}_{h\perp}^2$
and $Q^2=-q^2$ where $P$ is the target momentum.
The angles and other momenta are defined in the left panel of
Fig.~\ref{fig2-kinematics}, and $\varepsilon\sim (1-y)/(1-y+y^2/2)$ denotes
the polarization of the virtual photon with $y=(P\cdot q)/(P \cdot l)$, and
$S_T$ is the transverse spin vector. The dots in Eq.~(\ref{eq:structuref}) denote
terms with other angular distributions due to pretzelosity or subleading twist.

The parton model expressions for the structure functions in terms of convolutions
of TMD parton distributions and fragmentation functions can be easily understood
when using the "tree-level" formalism of Ref. \cite{Mulders:1995dh} where only
the leading order (in $\alpha_s$) tree diagrams in the hard part are considered.
By splitting the SIDIS cross section into a leptonic tensor $L_{\mu \nu}$
(assuming one-photon exchange) and a hadronic tensor $W^{\mu \nu}$,
$d\sigma_{\rm SIDIS} \propto L_{\mu \nu}W^{\mu \nu}$,
we can express the leading part (in $1/Q$) of $W^{\mu \nu}$ by the tree-diagram
in the right panel Fig.~\ref{fig2-kinematics}. The lower soft blob in that diagram
describes the momentum distribution of quarks inside the target whereas the upper
part represents the fragmentation of a parton into hadrons. Field theoretically the
soft blobs correspond to the following matrix elements of bilocal quark-quark
operators,
\begin{eqnarray}
\Phi_{ij}(x,\vec{p}_T^2|\eta) & = & \int \frac{dz^-d^2z_T}{(2 \pi)^3}\, \mathrm{e}^{ip\cdot z}\langle P,S\,|\,\bar{\psi}_j(0) \, \mathcal{W}_\Phi (0,z|\eta) \, \psi_i(z)\, |P,S\rangle \Big|_{z^+=0},\label{eq:Phi}\\
\Delta_{ij}(z,\vec{k}_T^2|\eta) & = & \frac{1}{2z}\sum_X \int \frac{dz^+ d^2z_T}{(2 \pi)^3} \, \mathrm{e}^{ik\cdot z}\langle 0| \mathcal{W}_\Delta [\infty,z|\eta]\, \psi_i(z)\,|P_h,X\rangle \times \nonumber\\
& & \hspace{2cm}\langle P_h,X|\,\bar{\psi}_j(0)\, \mathcal{W}_\Delta[0,\infty |\eta]\,|0\rangle \big|_{z^-=0}.\label{eq:Delta}
\end{eqnarray}
The gauge link operators $\mathcal{W}_\Phi$ and $\mathcal{W}_\Delta$ in Eqs.~(\ref{eq:Phi}) and (\ref{eq:Delta}) ensure the color gauge invariance of the matrix elements. They are defined as
\begin{eqnarray}
\mathcal{W}_\Phi[0,(z^-,0,\vec{z}_T)|\eta] & \equiv & [0\,|\,a_\Phi]\times [a_\Phi\,|\,b_\Phi]\times [b_\Phi\,|\,c_\Phi]\times  [c_\Phi \,|\,(z^-,0,\vec{z}_T)],\label{eq:Wilsonphi}\\
\mathcal{W}_\Delta[\infty,(0,z^+,\vec{z}_T)|\eta] & \equiv & \hspace{1cm}[b_\Delta \,|\,c_\Delta]\times [c_\Delta\,|\,(0,z^+,\vec{z}_T)]\,,\label{eq:WilsonDelta1}\\
\hspace{-0.3cm}\mathcal{W}_\Delta[0,\infty|\eta] & \equiv & \hspace{2cm}[0\,|\,a_\Delta]\times [a_\Delta\,|\,b_\Delta]\,,\label{eq:WilsonDelta2}
\end{eqnarray}
where [a|b] denotes a gauge link operator, i.e. a path-ordered exponential of gluon field operators, with a straight Wilson line between the space-time coordinates $a$ and $b$. The three "milestones" of the Wilson lines in Eqs.~(\ref{eq:Wilsonphi})-(\ref{eq:WilsonDelta2}) are $a_\Phi=(\eta\infty,0,\vec{0}_T)$, $b_\Phi=(\eta\infty,0,\vec{\infty}_T)$ and $c_\Phi=(\eta \infty,0,\vec{z}_T)$ for the correlator $\Phi$, while we have $a_\Delta=(0,\eta\infty,\vec{0}_T)$, $b_\Delta=(0,\eta\infty,\vec{\infty}_T)$ and $c_\Delta=(0,\eta\infty,\vec{z}_T)$ for the fragmentation correlator $\Delta$. Therein we use the usual light cone coordinates for a four vector $a^\mu=(a^-,a^+,\vec{a}_T)$ with $a^\pm=\frac{1}{\sqrt{2}}(a^0\pm a^3)$. The parameter $\eta \in \,\left\{-1,1\right\}$ determines the direction of the Wilson line. As was pointed out in \cite{Collins:2002kn} time-reversal transforms a Wilson line with $\eta=+1$ describing final state interactions, a situation one encounters in SIDIS, into a Wilson line with $\eta=-1$ describing initial state interaction, e.g. in Drell-Yan. This results in the relation (\ref{eq:Siversrel}) for the Sivers function.

It is convenient to express the diagram in the right panel of Fig.~\ref{fig2-kinematics} in terms of the correlators (\ref{eq:Phi}) and (\ref{eq:Delta}) in a frame in which the momentum of the nucleon $P$ and the momentum of the produced hadron $P_h$ are collinear. We obtain for the leading part in $1/Q$ of the hadronic tensor
\begin{eqnarray}
2MW^{\mu \nu} & = & 2z\sum_q e_q^2 \,\int d^2p_T\,d^2k_T\,\delta^{(2)}(\vec{p}_T+\vec{q}_T-\vec{k}_T)\,\times \nonumber\\
& & \hspace{4cm}\mathrm{Tr}\bigg[\Phi^q(x,\vec{p}_T)\gamma^\mu \Delta^q(z,\vec{k}_T)\gamma^\nu\bigg]+\mathcal{O}(1/Q).\label{eq:hadronicTensor}
\end{eqnarray}
The quark-quark correlators $\Phi$ and $\Delta$ contain all possible spin information of either the probed or fragmenting quark as well as the spin information of the target. It is possible to project out various polarizations of the quarks by tracing the correlators with appropriate Dirac matrices. These traces can be then parameterized in terms of TMD parton distributions and fragmentation functions \cite{Mulders:1995dh,Boer:1997nt} (for a complete parameterization see \cite{Goeke:2005hb}).
Assuming that the nucleon moves fast in one light cone direction, i.e. $P^+$ is the large component of the nucleon momentum, we obtain two TMD distributions for unpolarized quarks by tracing $\Phi$ with $\gamma^+$. One is the well-known ordinary PDF for an unpolarized target $f_1$ while the other is the Sivers function, describing the distribution of unpolarized quarks in a transversely polarized nucleon ($\epsilon_T^{ij}\equiv \epsilon^{-+ij}$ with $\epsilon^{0123}=+1$)
\begin{equation}
\frac{1}{2} \mathrm{Tr}\big[\Phi(x,\vec{p}_T)\gamma^+\big] \,=\, f_1(x,\vec{p}_T^2)-\frac{\epsilon_T^{ij}p_T^iS_T^j}{M}f_{1T}^\perp(x,\vec{p}_T^2)\label{eq:Phi+}.
\end{equation}
By tracing $\Phi$ with other Dirac structures we obtain further PDFs. For example,
by tracing it with $i\sigma^{i+}\gamma_5$ we obtain, among others, the transversity
distribution $h_1$ and the so-called 'pretzelosity' distribution $h_{1T}^\perp$, etc.

Since here we consider unpolarized hadrons in the final state only, there are only two relevant fragmentation functions, the ordinary fragmentation of unpolarized quarks and the chirally-odd Collins function $H_1^\perp$ of transversely polarized quarks. Assuming that the produced hadron moves fast in the minus light cone direction, i.e. the large component of $P_h$ is $P_h^-$, we project them out by tracing with $\gamma^-$ and $i\sigma^{i-\gamma_5}$,
\begin{eqnarray}
\frac{1}{2}\mathrm{Tr}\big[\Delta(z,\vec{k}_T)\gamma^-\big]\,=\,D_1(z,\vec{k}_T^2) & ; & \frac{1}{2}\mathrm{Tr}\big[\Delta(z,\vec{k}_T)i\sigma^{i-}\gamma_5\big]\,=\,-\frac{\epsilon^{ij}k_T^j}{M_h}H_1^\perp(z,\vec{k}_T^2).\label{eq:DeltaTr}
\end{eqnarray}

At this point one obtains the expression for the structure functions Sivers structure
function by inserting the traces of the type~(\ref{eq:Phi+}),~(\ref{eq:DeltaTr}) into
the hadronic tensor~(\ref{eq:hadronicTensor}), boosting into a frame in which the
momentum of the nucleon and of the virtual photon are collinear, and contracting
with the leptonic tensor $L_{\mu \nu}$ which yields:
\begin{eqnarray}
    \hspace{-1cm}
    F_{UT}^{\sin(\phi_h-\phi_s)}
    &=&     x \sum_q e_q^2 \int d^2p_T\,d^2k_T\,
        \delta^{(2)}(\vec{p}_T-\vec{k}_T-\vec{P}_{h\perp}/z)\nonumber\\
    &&  \hspace{3.5cm}
    \times \Big[-\frac{\vec{h}\cdot \vec{p}_T}{M_h}\,
    f_{1T}^{\perp,q}(x,\vec{p}_T^2)\,D_1^q(z,\vec{k}_T^2)\Big]\,,\label{Sivers}
\end{eqnarray}
where $\vec{h}=\vec{P}_{h\perp}/|\vec{P}_{h\perp}|$ is the normalized transverse
momentum of the produced hadron. The unpolarized structure function $F_{UU}$ is
given by (\ref{Sivers}) with
$(-\vec{h}\cdot \vec{p}_T)\,f_{1T}^{\perp,q}/M_h \to f_1^q$.

We emphasize again that the result (\ref{Sivers}) is valid for the tree-level hard
process. A factorization valid to all orders requires the inclusion of a soft factor
\cite{Ji:2004wu,Ji:2004xq,Collins:2004nx} in the formula (\ref{Sivers}) in order to
handle soft gluon radiation. Up to now no phenomenological treatment takes this
factor into account. We will neglect this factor in the following.

\section{SSA in $\mathbf{p^\uparrow p\to\pi X}$}
\label{Sec:Sivers-effect-in-hadron-hadron-collisions}

The first information on the Sivers function was obtained
from studies \cite{Anselmino:1994tv,D'Alesio:2007jt} of SSAs in $p^\uparrow p\to\pi X$
or $\overline{p}^\uparrow p\to\pi X$ \cite{Adams:1991rw,Adams:1991cs}.
Although they originally motivated the introduction of the Sivers
effect \cite{Sivers:1989cc,Sivers:1990fh} the theoretical understanding
of these processes is more involved compared to SIDIS or DY.
Here SSAs can also be generated by twist-3 effects
\cite{Efremov:1981sh,Efremov:1984ip,Qiu:1991pp,Qiu:1991wg,Qiu:1998ia},
though it was suggested that these could be manifestations of the same effect
in different $k_T$ regions \cite{Bomhof:2004aw,Ji:2006br,Ratcliffe:2007ye}.
Moreover, for this processes no factorization proof is formulated in terms
of the Sivers effect. Studies of other processes with $pp$ in the initial state
and hadronic final states indicate that it is in general hard to prove factorization
in hadronic reactions \cite{Bomhof:2004aw,Collins:2007nk,Collins:2007jp}.

\section{Sivers effect in SIDIS: first insights}
\label{Sec:Sivers-effect-in-SIDIS-first-insights}

The first information on the Sivers function from SIDIS was obtained
in \cite{Efremov:2004tp} from a study of preliminary HERMES data
\cite{Gregor:2005qv} on the 'weighted' SSA defined as
\be\label{Eq:SSA-in-SIDIS-weighted}
    A_{UT}^{P_{h\perp}/M_N \sin(\phi-\phi_S)}(x) \equiv \frac{1}{S_T}\,
    \frac{\sum_i\left\la\frac{P_{h\perp,i}}{M_N} N_i^\uparrow -
                        \frac{P_{h\perp,i}}{M_N} N_i^\downarrow\right\ra }
     {\sum_i\left\la\frac12(N_i^\uparrow + N_i^\downarrow)\right\ra}
    \ee
where $N_i^{\uparrow(\downarrow)}$ are sums over event counts for the respective
transverse target polarization, and $\la\dots\ra$ denotes averaging --- here
over $z$ and $P_{h\perp}$.
The advantage of 'weighted SSAs' is that the integrals in the structure
function (\ref{Sivers}) can be solved exactly
\cite{Boer:1997nt} yielding
\be\label{Eq:SSA-in-SIDIS-weighted-2}
    A_{UT}^{P_{h\perp}/M_N \sin(\phi-\phi_S)}(x,z) =
    \frac{2\int\di\vec{P}_{h\perp}^2 \frac{P_{h\perp}}{M_N}
    F_{UT}^{\sin(\phi-\phi_S)}(x,z,P_{h\perp})}
    {\int\di\vec{P}_{h\perp}^2 F_{UU}(x,z,P_{h\perp})}
    =\frac{(-2)\;
        \sum_a e_a^2\,x f_{1T}^{\perp(1)a}(x)\, D_1^{a}(z)}{
        \sum_a e_a^2\,x f_1^a(x)\,D_1^{a}(z)}
    \ee
where $f_{1T}^{\perp(1)a}(x) \equiv\int\!\di^2\vec{p}_T\frac{\vec{p}_T^2}{2 M_N^2}
f_{1T}^{\perp a}(x,\vec{p}_T^2)$.

\begin{figure}[t!]
\begin{tabular}{ccc}
\includegraphics[height=4cm]{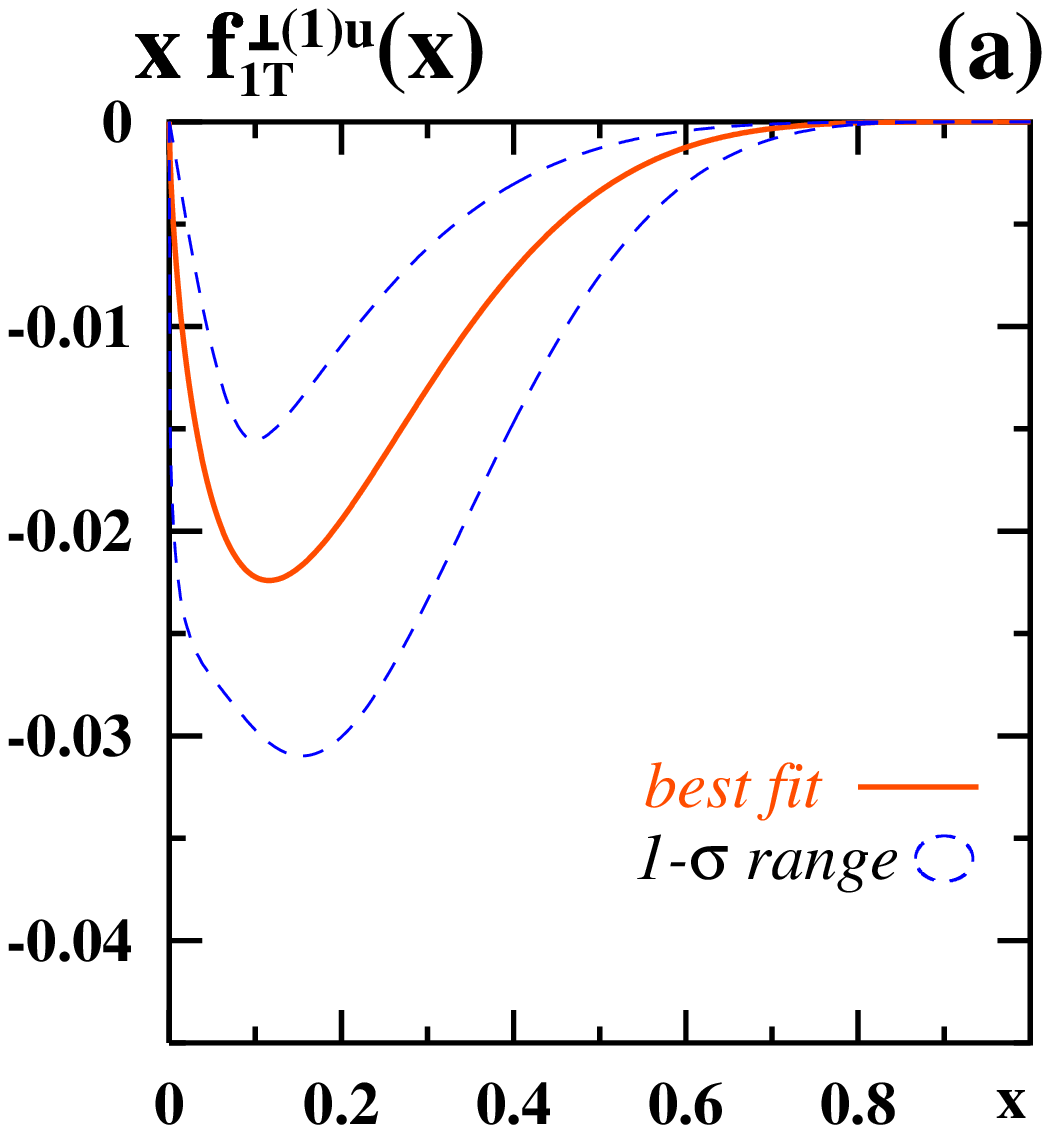}&
\includegraphics[height=4cm]{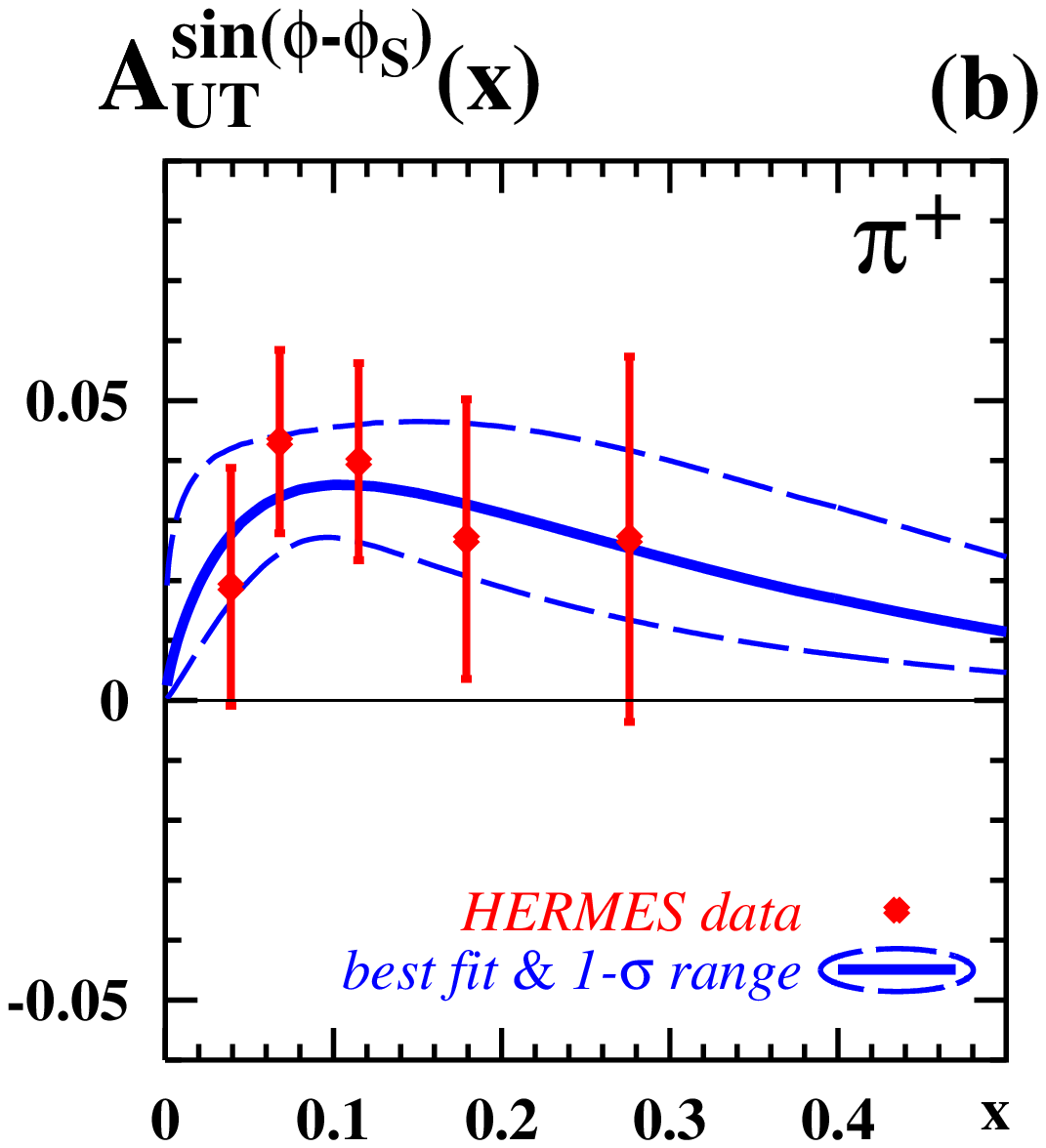}&
\includegraphics[height=4cm]{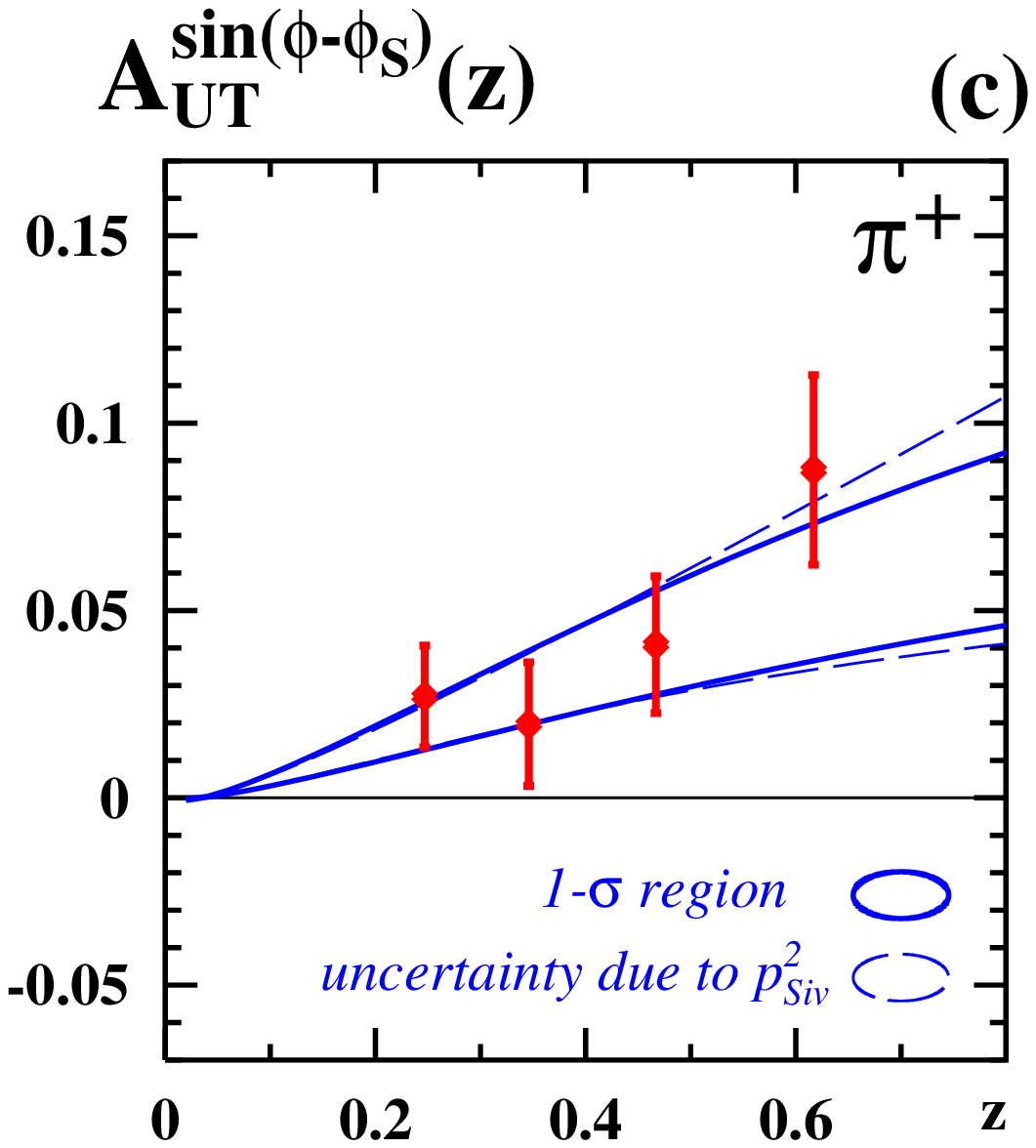}
\end{tabular}
\caption{\label{Fig5-compare-to-data}\footnotesize
        (a)
    Sivers function of $u$-quarks vs.\  $x$ at a scale of $2.5\,{\rm GeV}^2$,
    as obtained from HERMES data \cite{Airapetian:2004tw}. Shown are the best
        fit and its 1-$\sigma$ uncertainty.
        (b + c)
    The Sivers SSA,
    Eqs.~(\ref{Eq:SSA-in-SIDIS-unweighted},~\ref{Eq:AUT-SIDIS-Gauss}),
        for $\pi^+$ from proton as function of $x$ and $z$ as obtained from
    the fit in Figure~\ref{Fig5-compare-to-data}a in comparison to the data
    \cite{Airapetian:2004tw}. The $\pi^-$ data from \cite{Airapetian:2004tw}
    are compatible with zero, and are equally well described (not shown here).}
\end{figure}

While the weighting is preferable from a theory point of view,
it makes data analysis harder. It is difficult to control acceptance
effects, and the HERMES Collaboration
does not recommend the use of the preliminary data \cite{Gregor:2005qv}.
In 'unweighted SSAs' defined as
\be\label{Eq:SSA-in-SIDIS-unweighted}
    A_{UT}^{\sin(\phi-\phi_S)}(x) \equiv \frac{1}{S_T}\,
    \frac{\sum_i\left\la N_i^\uparrow - N_i^\downarrow \right\ra }
     {\sum_i\left\la\frac12( N_i^\uparrow + N_i^\downarrow)\right\ra}
    \ee
acceptance effects largely cancel. Therefore such data have been finalized
first, and one even is not discouraged to use preliminary data of this type
\cite{Diefenthaler:2005gx,Diefenthaler:2006vn,Diefenthaler:2007rj}.
However, the prize to pay is that now the convolution integrals in (\ref{Sivers})
can be solved only by resorting to models for the transverse momentum dependence.
Here we assume the distributions of transverse parton and hadron momenta in
distribution and fragmentation functions to be Gaussian with the corresponding
Gaussian widths, $p^2_{\rm Siv}$ and $K^2_{\!D_1}$, taken to be $x$- or
$z$- and flavor-independent. The Sivers SSA (\ref{Eq:SSA-in-SIDIS-unweighted})
as measured in
\cite{Airapetian:2004tw,Alexakhin:2005iw} is then given
by \cite{Collins:2005ie}
\be\label{Eq:AUT-SIDIS-Gauss}
        A_{UT}^{\sin(\phi-\phi_S)} = \frac{a_{\rm G}\,(-2)
        \sum_a e_a^2\,x f_{1T}^{\perp(1)a}(x)\, D_1^{a}(z)}{
        \sum_a e_a^2\,x f_1^a(x)\,D_1^{a}(z)}
        \;\;\;\mbox{with}\;\,\;
        a_{\rm G}=\frac{\sqrt{\pi}}{2}
        \frac{M_N}{\sqrt{p^2_{\rm Siv}+K^2_{\!D_1}/z^2}} \;.
    \ee
In view of the sizeable error bars of the first data it was necessary to
minimize the number of fit parameters. For that in \cite{Collins:2005ie}
effects of sea quarks were neglected. In addition, the prediction from
the limit of a large number of colors $N_c$ in QCD
\cite{Pobylitsa:2003ty}, namely
\be\label{Eq:large-Nc}
          f_{1T}^{\perp u}(x,\vec{p}_T^2) =
        - f_{1T}^{\perp d}(x,\vec{p}_T^2) \;\;\;
        \mbox{modulo $1/N_c$ corrections,} \ee
was imposed as an exact constraint. Analog relations holds also for antiquarks, and
all are valid for $x$ of the order $xN_c={\cal O}(N_c^0)$ \cite{Pobylitsa:2003ty}.
The following Ansatz was made and best fit obtained:
$       xf_{1T}^{\perp (1) u}(x) = -xf_{1T}^{\perp (1) d}(x)
        \stackrel{\rm Ansatz}{=}\;  A \, x^b   \,(1-x)^5
        \,\,\stackrel{\rm fit}{=}\,\, -0.17 x^{0.66}(1-x)^5$
\cite{Collins:2005ie,Collins:2005rq}.
Fig.~\ref{Fig5-compare-to-data}a shows
the fit and its 1-$\sigma$ uncertainty due to the
statistical error of the data \cite{Airapetian:2004tw}.
Fig.~\ref{Fig5-compare-to-data}b shows that this fit well
describes the $x$-dependence of the HERMES data \cite{Airapetian:2004tw}.
Fig.~\ref{Fig5-compare-to-data}c finally shows the equally good
description of the $z$-dependence of the data \cite{Airapetian:2004tw},
that were not included in the fit. This serves as a cross check
for the Gauss Ansatz, which apparently works well here.
It is found in general that this model is useful as long
as one deals with limited precision and small transverse momenta
$\la P_{h\perp}\ra \ll Q$ \cite{D'Alesio:2004up,D'Alesio:2007jt}.

%
%

In \cite{Collins:2005ie} it was furthermore found that effects due to sea quarks
could not be resolved within the error bars of the data \cite{Airapetian:2004tw}.
It was also checked that $1/N_c$-corrections expected in (\ref{Eq:large-Nc})
are within the error bars of HERMES \cite{Airapetian:2004tw} and especially
COMPASS \cite{Ageev:2006da} data.

To draw an intermediate conclusion, the HERMES and COMPASS data
\cite{Airapetian:2004tw,Alexakhin:2005iw} are compatible with
large-$N_c$ predictions \cite{Pobylitsa:2003ty}.
More precisely, the large-$N_c$ approach worked {\sl at that stage}, because
the precision of the first data \cite{Airapetian:2004tw,Alexakhin:2005iw} was
comparable to the theoretical accuracy of the large-$N_c$ relation
(\ref{Eq:large-Nc}). Remarkably, the signs of the extracted Sivers functions,
$f_{1T}^{\perp u} < 0$ and  $f_{1T}^{\perp d} > 0$,  agree with the physical
picture discussed in \cite{Burkardt:2002ks}.
The findings of \cite{Collins:2005ie} were in agreement with
other studies \cite{Anselmino:2005ea,Vogelsang:2005cs},
see also the review \cite{Anselmino:2005an}.

\section{Sivers effect in DY: First predictions}
\label{Sec:Sivers-effect-in-DY}

As the experimental test of the particular 'universality relation' for the
Sivers function in Eq.~(\ref{eq:Siversrel}) is of fundamental importance,
the first insights on $f_{1T}^\perp$ \cite{Efremov:2004tp,Collins:2005ie}
were immediately used to estimate the feasibility of DY-experiments to
measure the Sivers effect \cite{Efremov:2004tp,Collins:2005ie}.
It was shown that the Sivers effect leads sizeable SSAs in
$p^\uparrow\pi^-\to l^+l^-X$, which can be measured at COMPASS, and in
$p^\uparrow\bar{p}$ or $p\bar{p}^\uparrow\to l^+l^-X$, which could
be studied in the proposed PAX experiment at FAIR, GSI \cite{Barone:2005pu}.

On a shorter term the Sivers effect in DY can be studied in
$p^\uparrow p\to l^+l^-X$ at RHIC. In $pp$-collisions sea quarks are involved,
and counting rates are smaller. It was shown, however, that the Sivers SSA in DY
can nevertheless be measured at RHIC with an accuracy sufficient to unambiguously
test Eq.~(\ref{eq:Siversrel}) for Sivers quarks \cite{Collins:2005ie,Collins:2005wb}.
Moreover, RHIC can provide valuable information on Sivers antiquarks
\cite{Collins:2005ie,Collins:2005wb}.

\section{ Sivers effect in SIDIS: Further developments}
\label{Sec:Sivers-effect-in-SIDIS-further-developments}

The first data \cite{Airapetian:2004tw,Alexakhin:2005iw} gave rise to a
certain optimism. The Sivers effect on a proton target was clearly seen
\cite{Airapetian:2004tw}, its smallness on a deuteron target \cite{Alexakhin:2005iw}
understood in theory \cite{Pobylitsa:2003ty}. Independent studies
\cite{Efremov:2004tp,Anselmino:2005ea,Vogelsang:2005cs,Collins:2005ie} agreed on
the interpretation of the SIDIS data \cite{Anselmino:2005an}.
Prospects of accessing the effect in DY (and to test the universality
property of $f_{1T}^\perp$) were found promising, see
\cite{Efremov:2004tp,Anselmino:2005ea,Vogelsang:2005cs,Collins:2005rq}
and \cite{Bianconi:2005yj,Bianconi:2006hc,Radici:2007vc}.
The optimism persisted the following releases of higher statistics
data \cite{Diefenthaler:2005gx,Ageev:2006da} from HERMES and COMPASS,
but it was somehow damped with the advent of HERMES data on kaon Sivers SSAs
\cite{Diefenthaler:2006vn,Diefenthaler:2007rj}, while COMPASS reconfirmed
that on deuteron the effect is really small
\cite{Martin:2007au,Vossen:2007mh,Alekseev:2008dn}.

Let us first discuss what one would naively expect for the $K^+$
Sivers effect. Comparing the 'valence quark structure' ($K^+ \sim
u\overline{s}$ vs.\ $\pi^+\sim  u\overline{d}$), we see that the
DIS-production of $K^+$ and $\pi^+$ differs by the 'exchange' of
the sea quarks $\overline{s}\leftrightarrow \overline{d}$. If it
were legitimate to neglect sea quarks in nucleon, then the
$K^+$ and $\pi^+$ Sivers SSAs would be comparable.

\begin{figure}[t!]
\begin{tabular}{ccc}
\includegraphics[height=4cm]{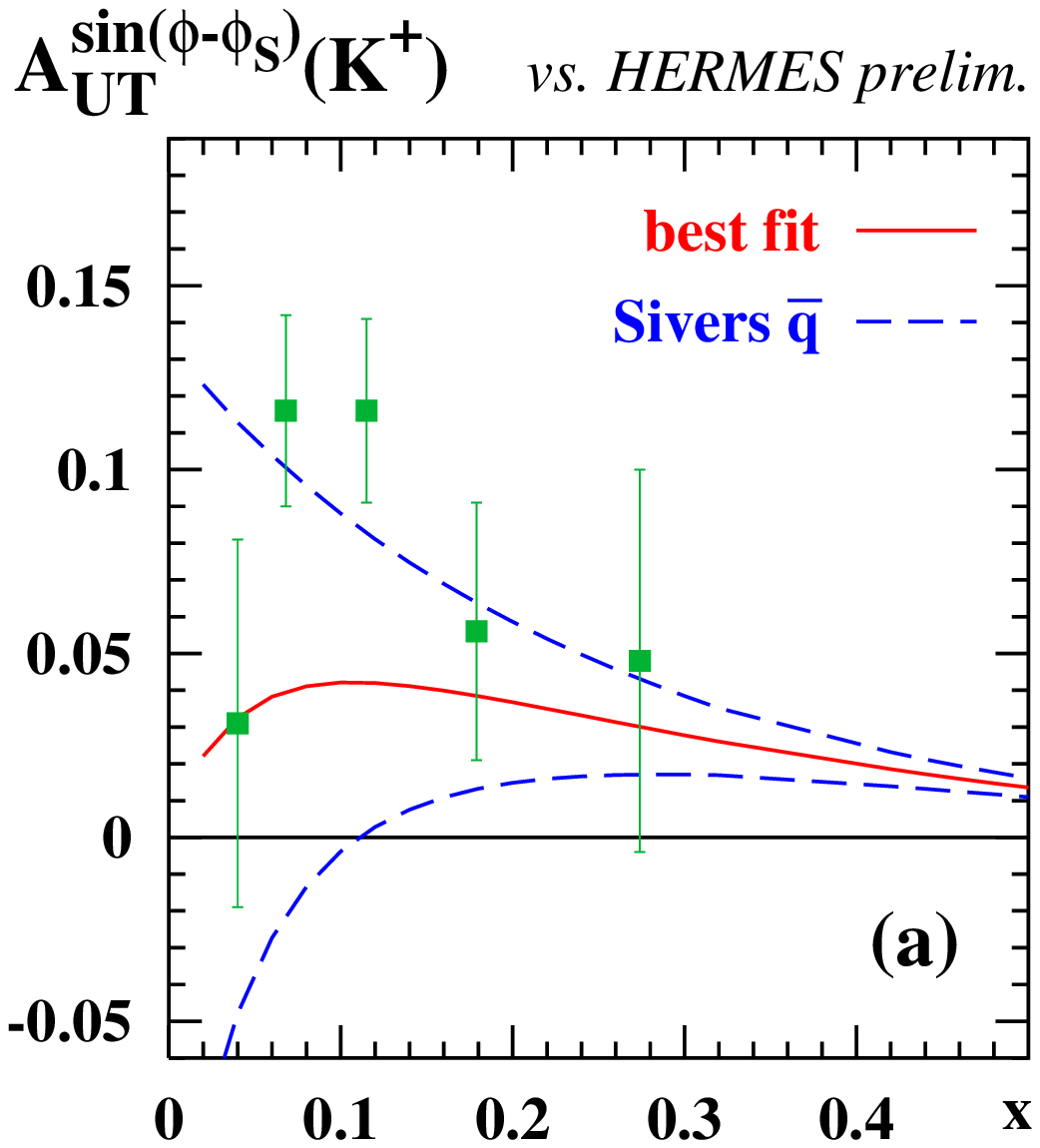} &
\includegraphics[height=4cm]{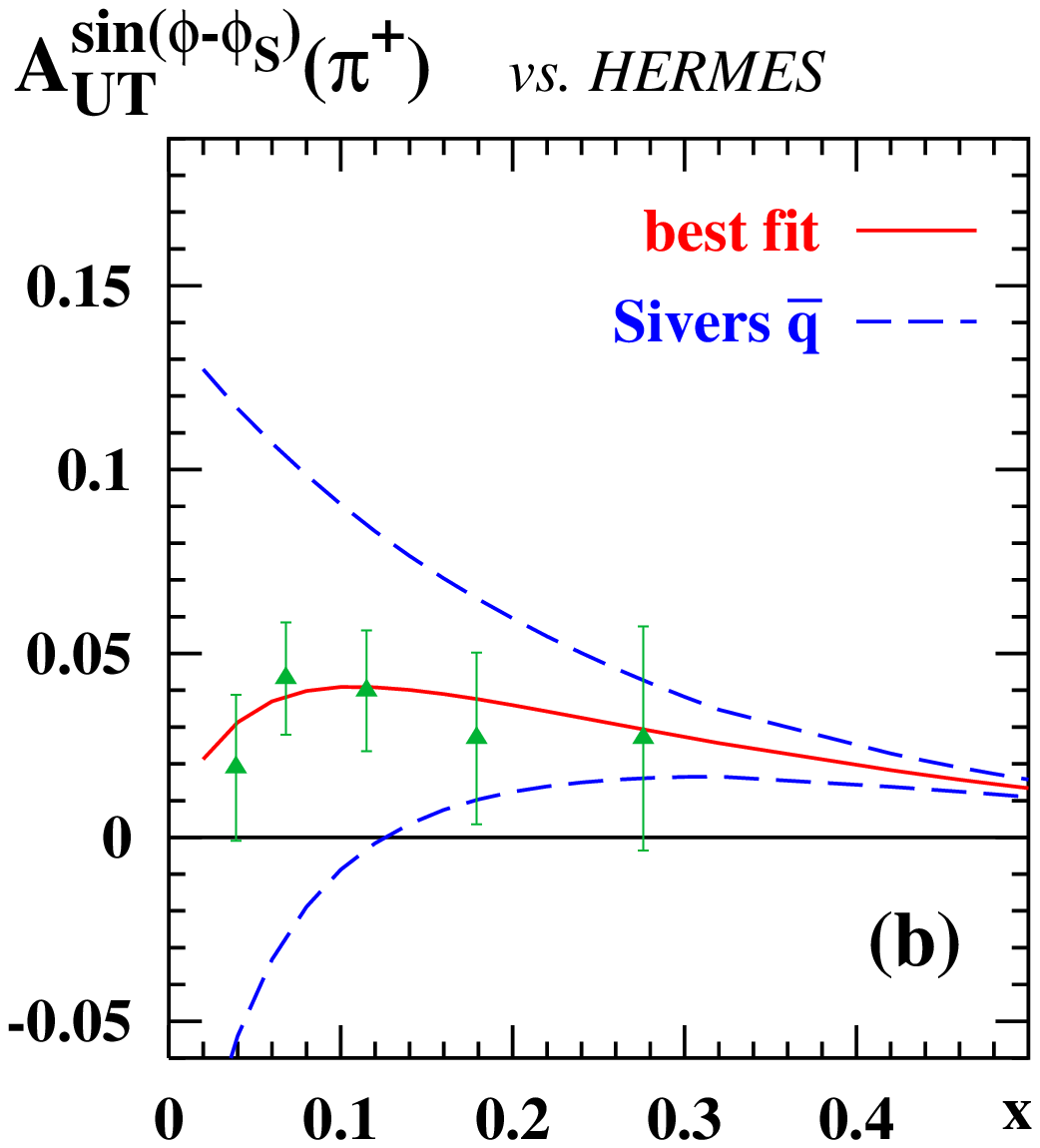} &
\includegraphics[height=4cm]{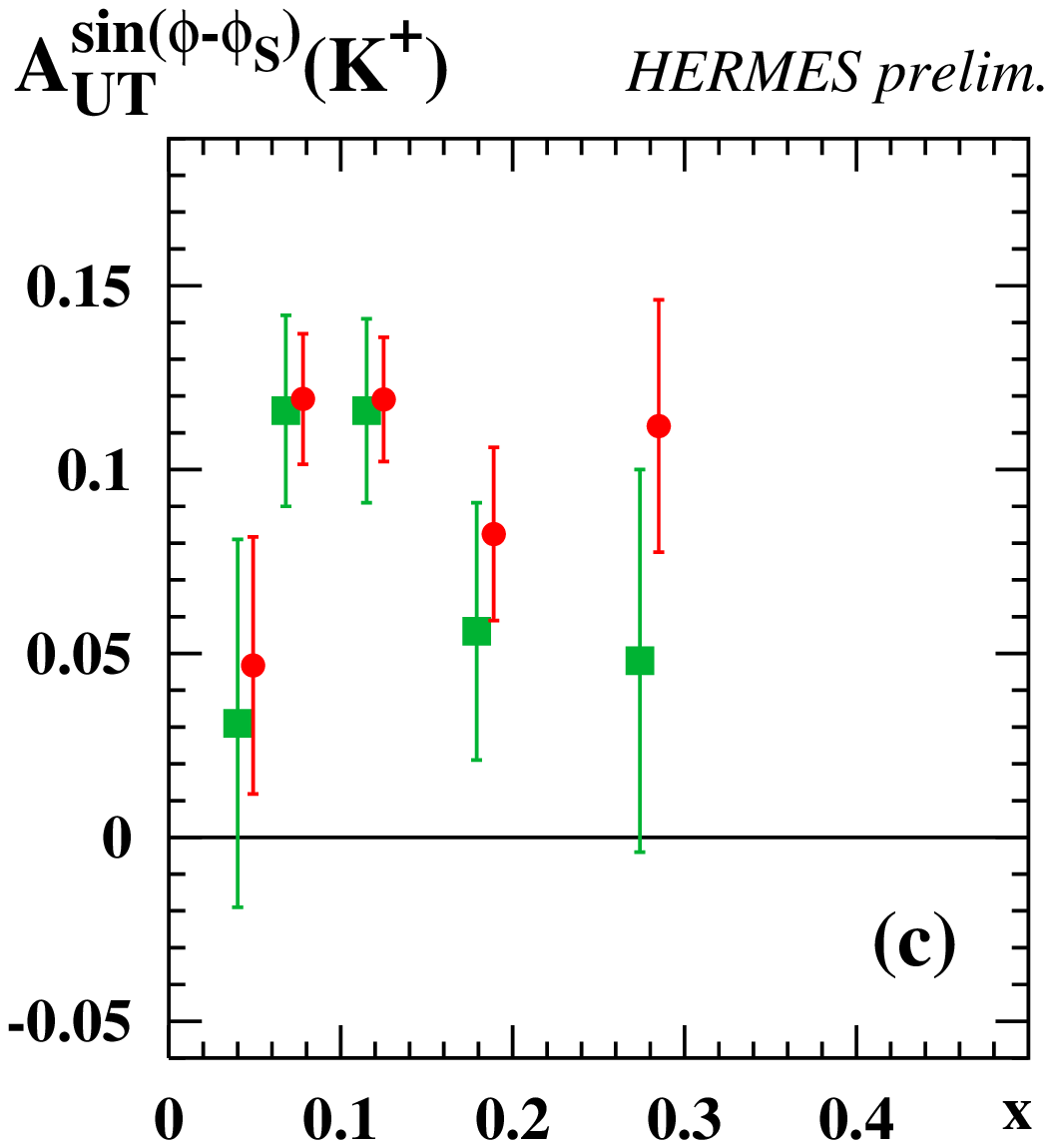}
\end{tabular}
\caption{\label{Fig4-new-kaon-data}\footnotesize
        (a)
    The Sivers SSA for $K^+$ as function of $x$. The preliminary HERMES data is
    from \cite{Diefenthaler:2006vn}. The solid line is the $K^+$ SSA obtained
    from the best fit to pion data \cite{Airapetian:2004tw}
    (see Sec.~\ref{Sec:Sivers-effect-in-SIDIS-first-insights}).
    The dashed lines display the effect of adding on top of that
    Sivers sea quarks saturating $\pm$ the positivity bounds
    (see Sec.~\ref{Sec:Sivers-effect-in-SIDIS-further-developments}).
    It seems that sea quarks could explain the effect, but at the same
    time one overshoots $\pi^+$ data (see next figure).
    (b)~Sivers SSA for $\pi^+$ as function of $x$.
    The published HERMES data are from \cite{Airapetian:2004tw}, and the
    theoretical curves as in Fig.~\ref{Fig4-new-kaon-data}b (but for $\pi^+$).
    (c)
    Comparison of the first (lower statistics, boxes) \cite{Diefenthaler:2006vn}
    and the most recent (higher statistics, circles) \cite{Diefenthaler:2007rj}
    data from HERMES on the $K^+$ Sivers SSA as function of~$x$. It seems
    unlikely that the effect could be due to statistical fluctuation,
    especially in the region of $x\sim 0.1$.}
\end{figure}

Now let us have a look on the preliminary HERMES data \cite{Diefenthaler:2006vn}.
At larger $x > 0.15$ the Sivers SSAs for $K^+$ and $\pi^+$ are comparable, as expected.
The fit to pion data discussed in Sec.~\ref{Sec:Sivers-effect-in-SIDIS-first-insights}
well describes the $K^+$ data in this region \cite{Efremov:2007kj,Efremov:2008vf},
see the solid line in Fig.~\ref{Fig4-new-kaon-data}a
(where, of course, all 'differences' between kaons and pions due to
different fragmentation functions \cite{Kretzer:2000yf} are considered).
However, at smaller $x\sim 0.1$ we observe a (2-3) times larger SSA for $K^+$
compared to $\pi^+$. For $K^-\sim s\overline{u}$ (i.e.\ pure 'sea quark effect')
the SSA is compatible with zero and bears no surprises.

The $K^+$ Sivers effect at HERMES \cite{Diefenthaler:2006vn,Diefenthaler:2007rj}
hints at an importance of Sivers sea quarks.
Interestingly, no kaon over pion enhancement is observed at COMPASS
\cite{Martin:2007au,Vossen:2007mh,Alekseev:2008dn}. This is remarkable, because
the COMPASS kinematics covers much smaller $x$ down to $x_{\rm min}=0.003$
(vs.\ HERMES $x_{\rm min}=0.023$). Sea quark effects could therefore
show up at COMPASS even more clearly --- however, one has to keep in mind
the different targets: proton at HERMES vs.\ deuteron at COMPASS.

In order to get a feeling about the impact of sea quark effects,
let us do the following exercise. We use for $f_{1T}^{\perp u}$ and
$f_{1T}^{\perp d}$ the previous best fit results,
see Sec.~\ref{Sec:Sivers-effect-in-SIDIS-first-insights}, and add
on top of that Sivers $\bar u$, $\bar d$, $s$ and $\bar s$-distributions
which saturate the positivity bounds \cite{Bacchetta:1999kz}
\be\label{Eq:Sivers-positivity}
    |f_{1T}^{\perp(1) a}(x)| \le \frac{\la p_T^{\,a}\ra_{\rm unp}}{2M_N} f_1^a(x) \;,
\ee
where $\la p_T^{\,a}\ra_{\rm unp}$ is the mean transverse momentum
of unpolarized quarks in the nucleon \cite{Collins:2005ie}.
(It could depend on flavor. Neglecting this possibility one obtains a good
des\-cription of transverse hadron momenta at HERMES \cite{Airapetian:2002mf} for
$\la p_T\ra_{\rm unp} = 0.5\,{\rm GeV}$ \cite{Collins:2005ie}.)
The effects of sea quarks allowed to saturate $\pm$ these bounds are shown in
Fig.~\ref{Fig4-new-kaon-data}a as dashed lines. Throughout the parameterization
\cite{Gluck:1998xa} is used for $f_1^a$.

Fig.~\ref{Fig4-new-kaon-data}a demonstrates that sea quarks may
have a strong impact, and could be able to explain the Sivers $K^+$ SSA.
However, introducing Sivers sea quarks as large as to explain the $K^+$ data,
Fig.~\ref{Fig4-new-kaon-data}a, tends to overshoot the $\pi^+$ data, see
Fig.~\ref{Fig4-new-kaon-data}b. Also there is no reason to expect
Sivers sea quarks to saturate positivity bounds.

If the Sivers $K^+$ effect is not a statistical fluctuation
(which is unlikely, see Fig.~\ref{Fig4-new-kaon-data}c),
then it should be possible to obtain a satisfactory fit
to the pion and kaon data from HERMES and COMPASS.
The next Section is devoted to this task.

%
%

\section{  Understanding pion and kaon Sivers effect}
\label{Sec:Understanding-pion-and-kaon-Sivers-effect}

In order see whether it is possible to understand the data on the Sivers SSAs
for pions and kaons from different targets,
it is necessary to attempt a simultaneous fit. We use the
HERMES proton target data on $\pi^0$, $\pi^\pm$, $K^\pm$
from \cite{Diefenthaler:2007rj} and the COMPASS deuteron data
on $\pi^\pm$, $K^\pm$ from \cite{Martin:2007au}.
Since we do not know error correlation matrices
we cannot perform a simultaneous fit to data on $x$, $z$ and
$P_{h\perp}$-dependences with correct estimate of its statistical
significance. In the present study data on $x$-dependence will serve
as input for the fit, and data  on $z$-dependence will be used
only for a cross check of the results.
When using data from different experiments it is necessary to consider
systematic uncertainties. Those are dominated by the uncertainty of the
target polarization in both experiments, and we combine them with
statistical errors in quadrature.

The data on $x$-dependences of the Sivers SSAs have little sensitivity to the
parameters entering the factor $a_{\rm G}$ in (\ref{Eq:AUT-SIDIS-Gauss}).
We fix the Gaussian width of the fragmentation function and Sivers function
as $\la K^2_{D_1}\ra = 0.16\,{\rm GeV}^2$ \cite{Collins:2005ie},
and $\la p^2_{\rm Siv}\ra=0.2\,{\rm GeV}^2$
which coincides with the central value obtained in \cite{Collins:2005ie}.
We stress that the final results presented here anyway depend only weakly on
these parameters, which could be inferred from studies of data on
$P_{h\perp}$-dependence. Such studies will be reported elsewhere. Finally,
for $f_1^a(x)$ and $D_1^a(z)$ we use the leading order parameterizations
\cite{Kretzer:2000yf,Gluck:1998xa} at a scale of $2.5\,{\rm GeV}^2$
which is close to the $\la Q^2\ra$ of both experiments.

%
\begin{figure}
\begin{tabular}{cc}
\includegraphics[height=3.8cm]{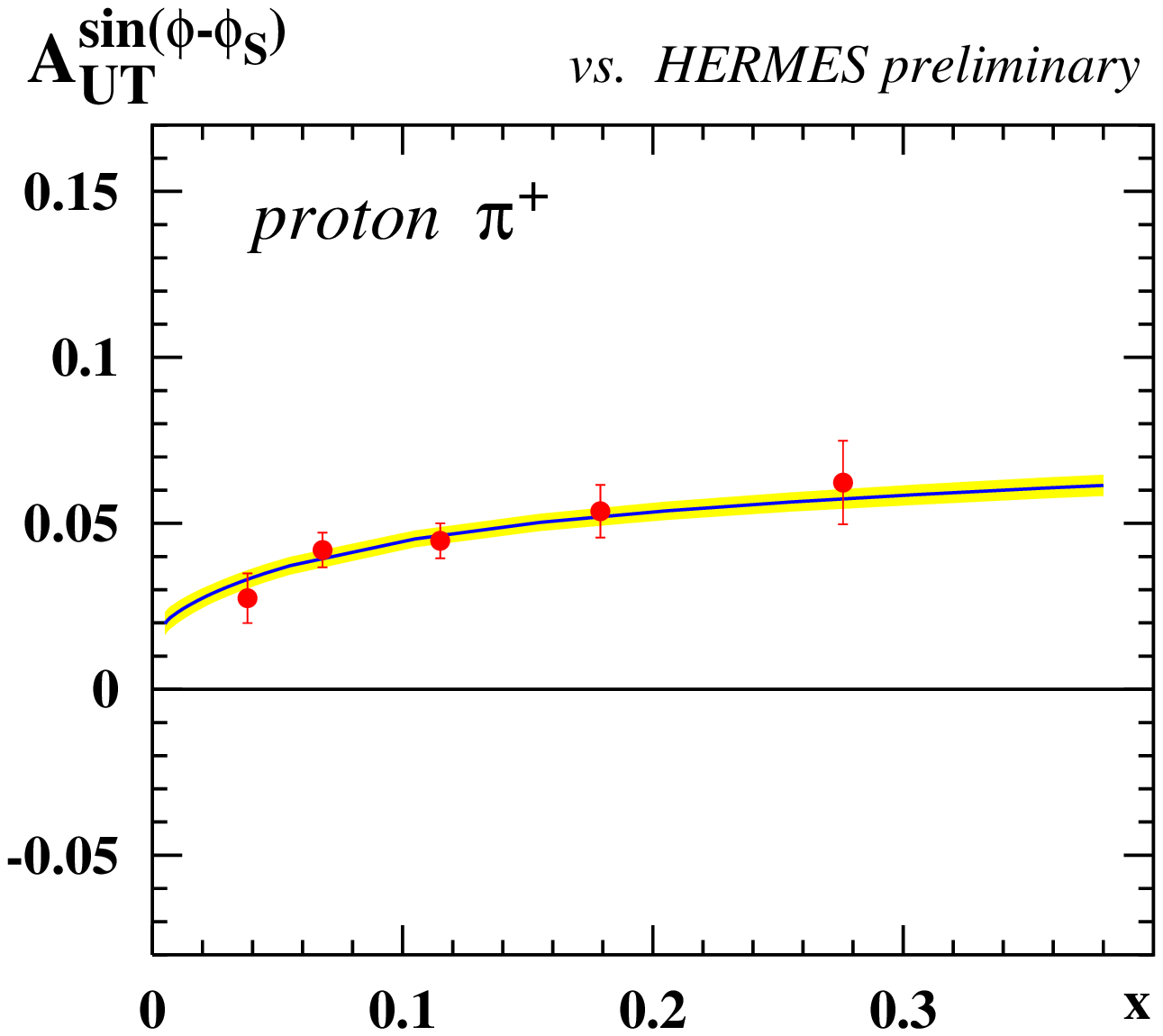}&
\includegraphics[height=3.8cm]{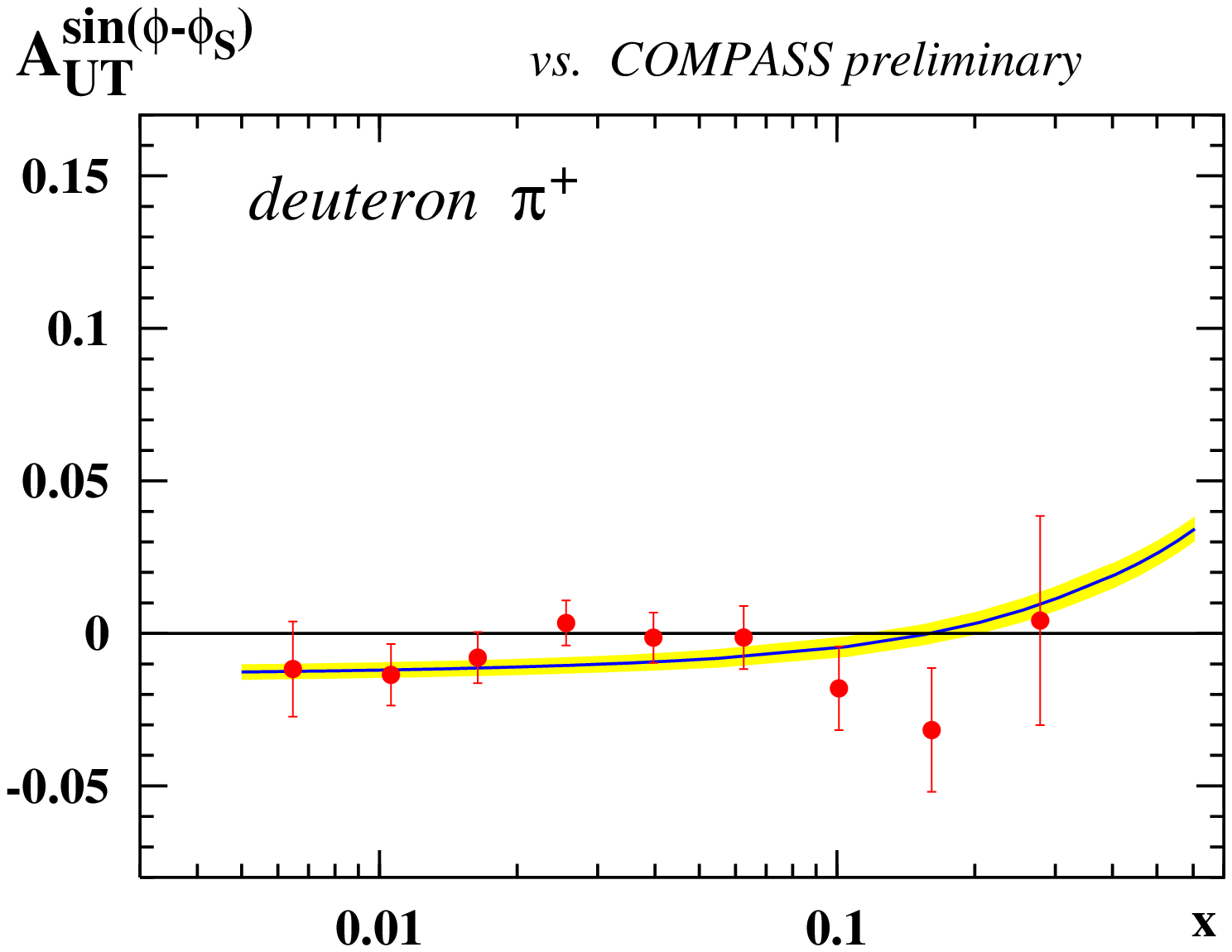}\\
\includegraphics[height=3.8cm]{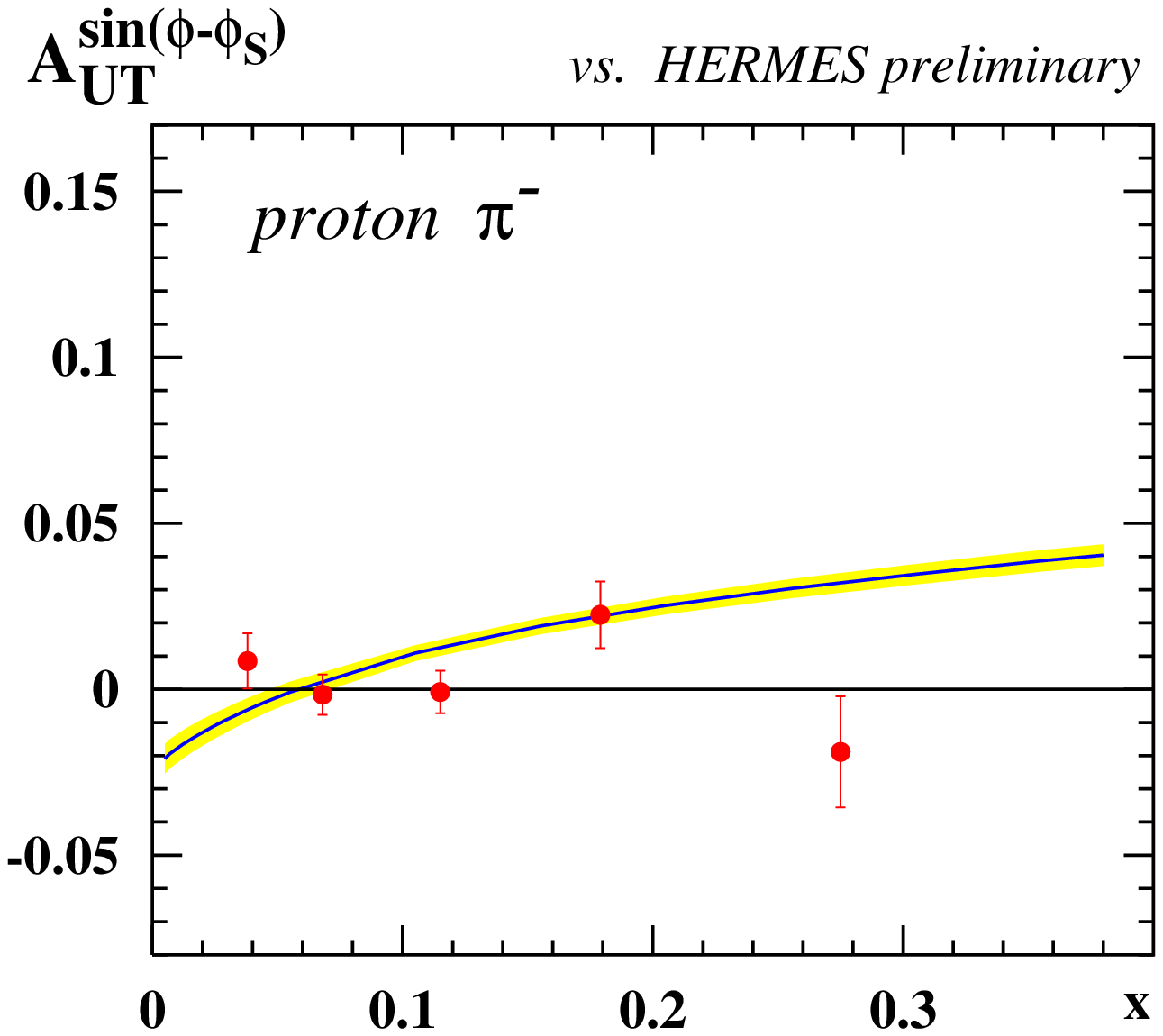}&
\includegraphics[height=3.8cm]{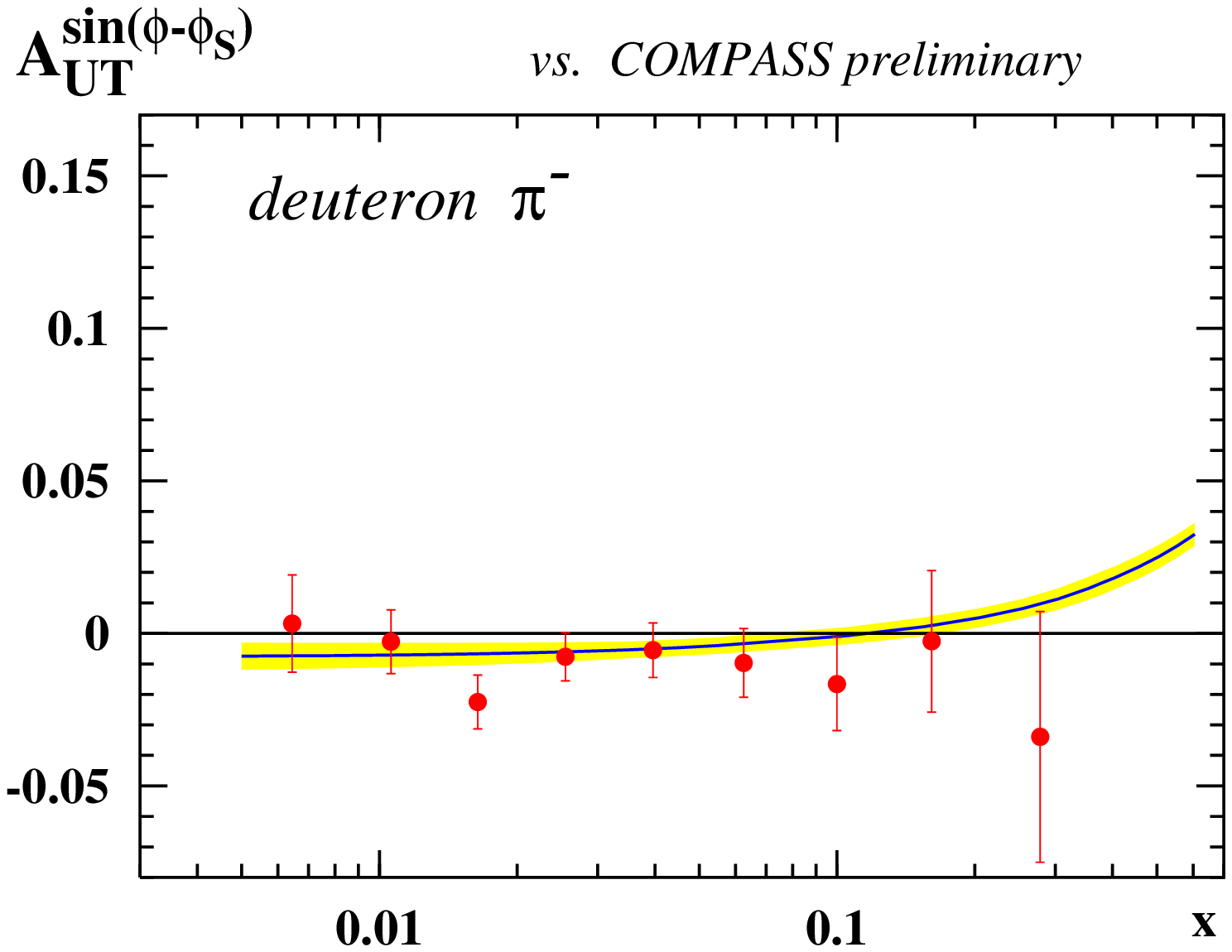}\\
\includegraphics[height=3.8cm]{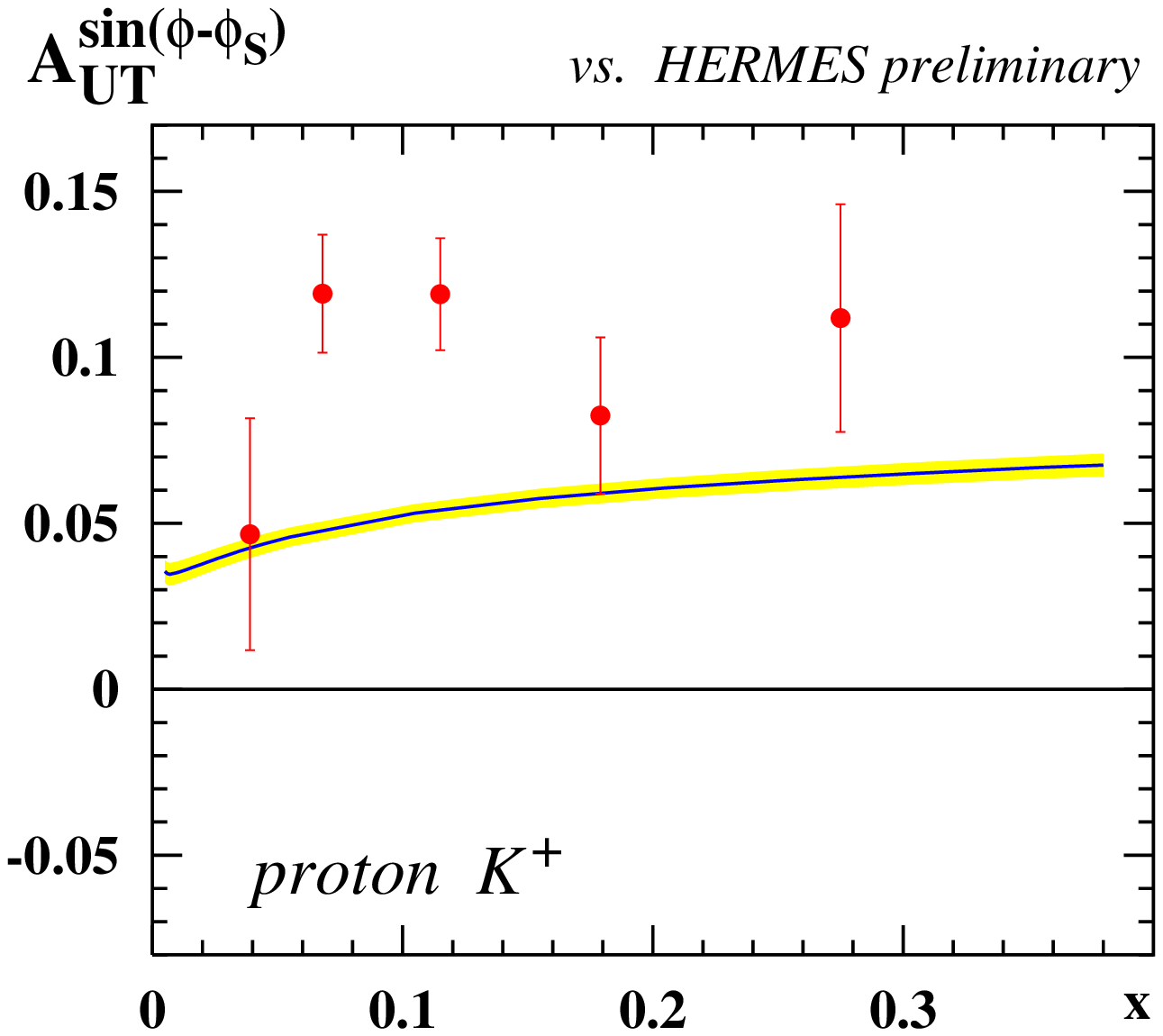}&
\includegraphics[height=3.8cm]{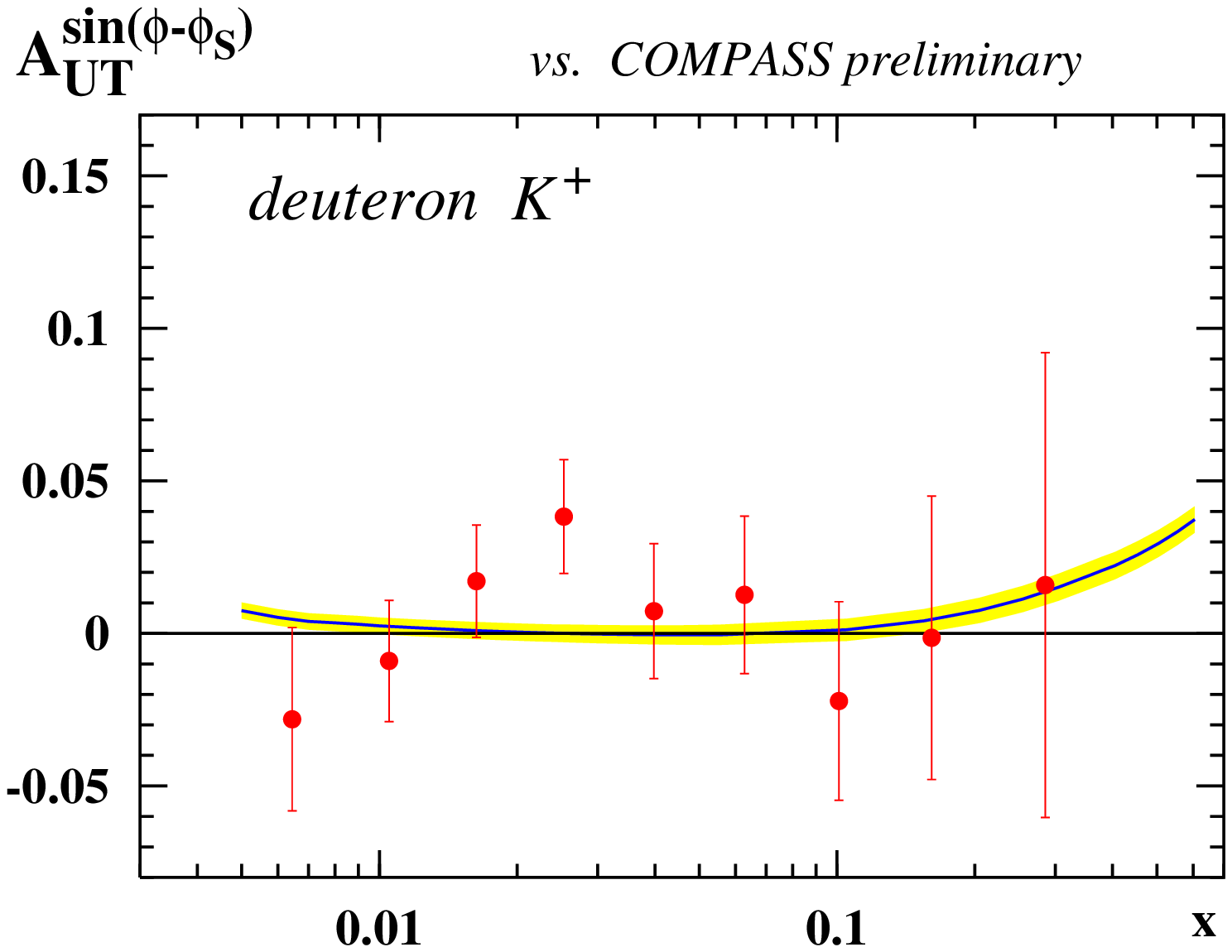}\\
\includegraphics[height=3.8cm]{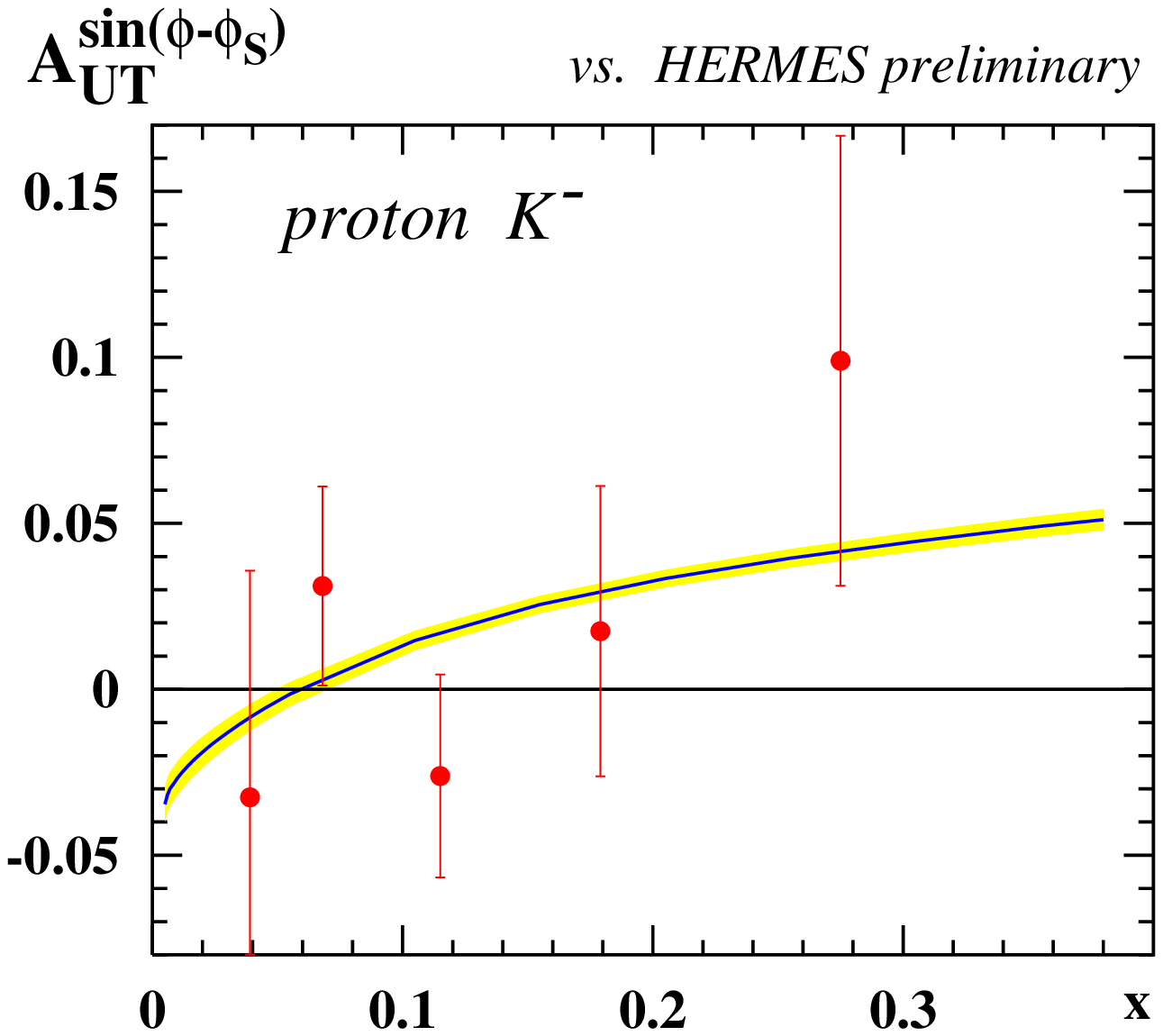}&
\includegraphics[height=3.8cm]{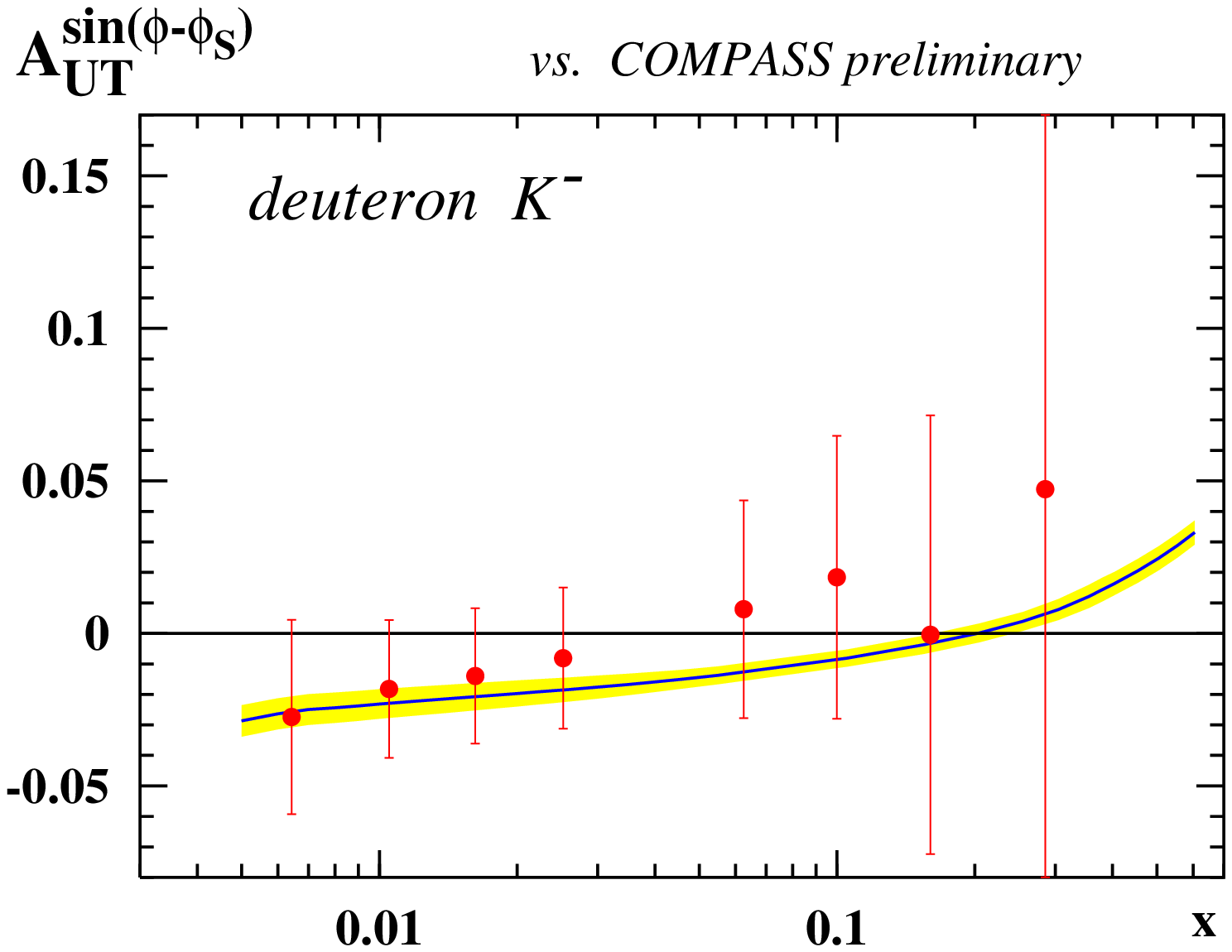}\\
\includegraphics[height=3.8cm]{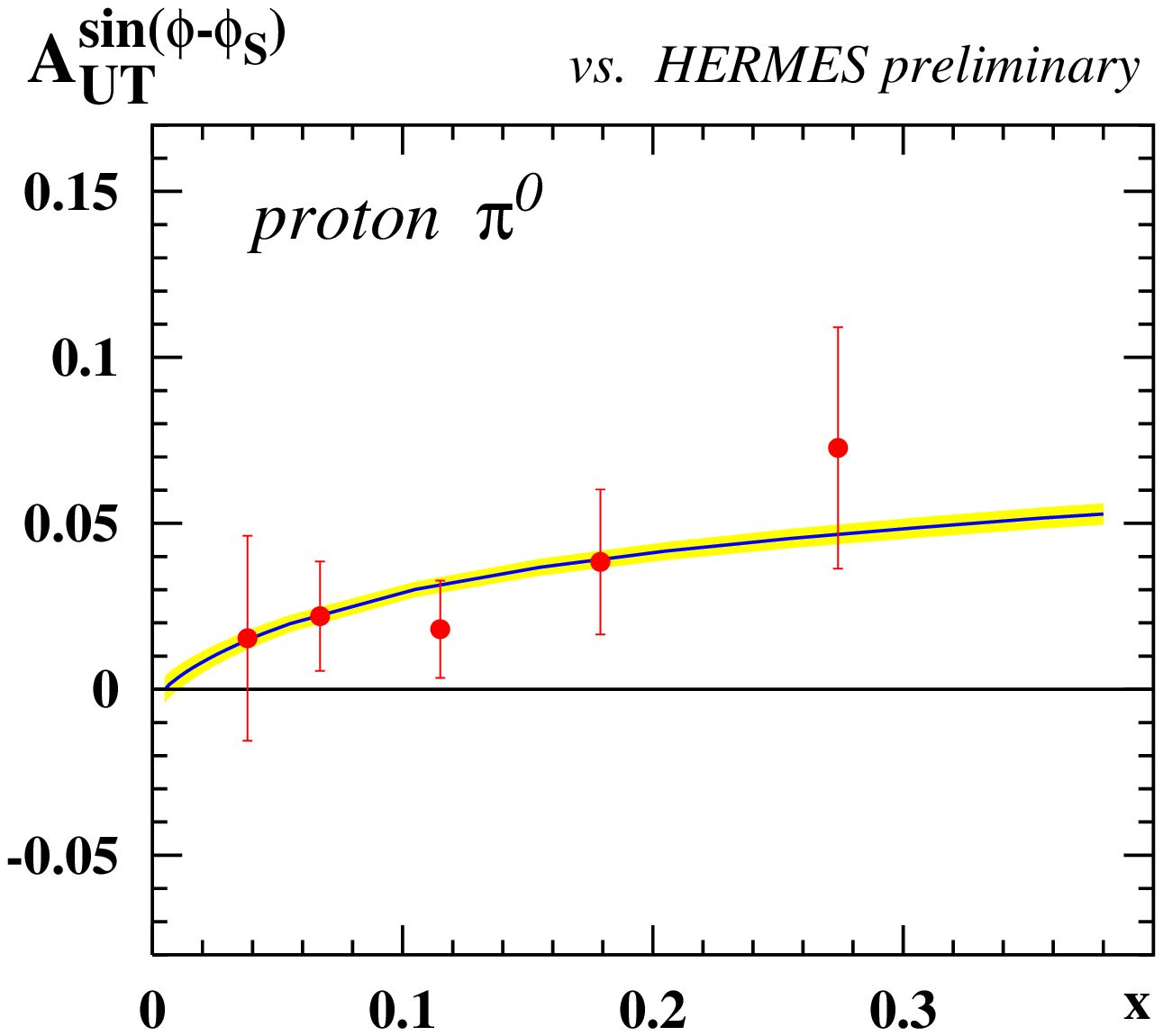}&
\end{tabular}
\caption{\label{Fig5:AUT-Siv-x}
    The Sivers SSA for various hadrons from different targets vs.\ $x$.
    Left  panel: HERMES data (proton target) \cite{Diefenthaler:2007rj}.
    Right panel: COMPASS data (deuteron target) \cite{Martin:2007au}.
    The theoretical curves are the best fit (solid line) and its
    1-$\sigma$-region (shaded area) as obtained from the Ansatz
    and best fit in Eqs.~(\ref{Eq:f1Tperp-ansatz},~\ref{Eq:best-fit}).
    These data served as INPUT for the fit
    (\ref{Eq:f1Tperp-ansatz},~\ref{Eq:best-fit}).}
\end{figure}
%

Now we need an Ansatz for the Sivers functions.
Here we shall content ourselves to the following simple Ansatz (the
constraints on the $A_a$ arise from positivity, Eq.~(\ref{Eq:Sivers-positivity}))
\be\label{Eq:f1Tperp-ansatz}
    f_{1T}^{\perp(1)a} = A_a \; \frac{\la p_T \ra_{\rm unp}}{2M_N}\,f_1^a(x)\;,
    \;\;\;|A_a|\le 1
\ee
with $\la p_T \ra_{\rm unp}=0.5\,{\rm GeV}$ from \cite{Collins:2005ie}.
Since sea quark effects are of importance (see
Sec.~\ref{Sec:Sivers-effect-in-SIDIS-further-developments}) we introduce
the flavours: $a=u,\,d,\,s,\,\overline{u},\,\overline{d},\,\overline{s}$.
So the initial task is to fix the $N_{\rm para} = 6$ parameters $A_a$
from $N_{\rm data}=N_{\rm HERMES} + N_{\rm COMPASS} = 25 + 36$ data points.
The Ansatz (\ref{Eq:f1Tperp-ansatz}) makes the numerical handling of the problem
particularly simple. The $\chi^2$ is a 'six-dimensional parabola' in the space of
the $A_a$, and the only extremum (global minimum)  is easily found by means
of simple self-made codes, or {\tt minuit} \cite{James:2006}.

The best fit has a $\chi^2=73.4$.
This means a satisfactory $\chi^2$ per degree of freedom of
$\chi^2/(N_{\rm data}-N_{\rm para})\equiv \chi^2_{\rm d.o.f.}=1.33$.
The results for the best fit parameters read:
$A_u        = -0.21$,
$A_d        =  0.41$,
$A_{\bar u} =  0.24$,
$A_{\bar d} = -0.27$,
$A_s        =  0.95$,
$A_{\bar s} = -1.93$.
The (correlated!) 1-$\sigma$ uncertainties of these fit results are of
${\cal O}(10\%)$.
But $A_{\bar s}$ exceeds the positivity bound (\ref{Eq:f1Tperp-ansatz}) and $A_s$
comes suspiciously close to it, which is driven by the $K^+$ HERMES data.

We therefore repeat the fit and use the above results to inspire the following
Ansatz. We fix
$A_s        = +1$ and
$A_{\bar s} = -1$
from the very beginning. This 4-parameter-fit has, of course, a slightly higher
$\chi^2= 76.5$ but nearly the same $\chi^2_{\rm d.o.f.}=1.34$ which means that
it is equally good. The best fit parameters read
\ba
A_u       =         - 0.21\pm 0.01 &
A_{\bar u}=\phantom{-}0.23\pm 0.02 &A_s       \stackrel{\rm fixed}{=}+1\nonumber\\
A_d       =\phantom{-}0.38\pm 0.03 &
A_{\bar d}=         - 0.28\pm 0.04 &A_{\bar s}\stackrel{\rm fixed}{=}-1.
    \label{Eq:best-fit}\ea
The 1-$\sigma$ errors are those solutions for the $A_a$
($a=u,\,d,\,\overline{u},\,\overline{d}$) which increase the total
$\chi^2 = 76.5$ by one unit. For our Ansatz this is again particularly simple.
For example, the uncertainty of $A_u$ is found by fixing the other free parameters
to their best fit values, and solving a quadratic equation.
The  1-$\sigma$ uncertainties in (\ref{Eq:best-fit}) are correlated, of course.

Fig.~\ref{Fig5:AUT-Siv-x} shows how the fit
(\ref{Eq:f1Tperp-ansatz},~\ref{Eq:best-fit}) describes the data on the
$x$-dependence of the Sivers SSAs (i.e.\ the input for the fit).
Let us comment on the fit:
\begin{itemize}
\item   The fit quality is satisfactory: $\chi^2_{\rm d.o.f.}=1.3={\cal O}(1)$
	as it should be.
	(There are stronger criteria for goodness of a fit \cite{Collins:2001wx}, 
	but we have too few data to apply them.)
\item   Looking at Fig.~\ref{Fig5:AUT-Siv-x} we notice: out of 61 data points,
        there are only two(!) data points that are off the best fit curve
        in a worthwhile mentioning way.
\item   From the point of view of statistics, one can comfortably live with such
        a situation, and wait for new data that will allow to improve the fit.
\item   But is it not suspicious that those two points, that are off the best fit
        in a worthwhile mentioning way, are precisely the $K^+$ HERMES data
        around $x\sim0.1$?
\end{itemize}
The last observation raises the question, whether the $K^+$ HERMES data around
$x\sim0.1$ could be a statistical fluctuation. Of course, this possibility
cannot be excluded --- though one is not inspired to consider such an explanation
as convincing, looking back at Fig.~\ref{Fig4-new-kaon-data}c.

%
\begin{figure}
\begin{tabular}{cc}
\includegraphics[height=3.8cm]{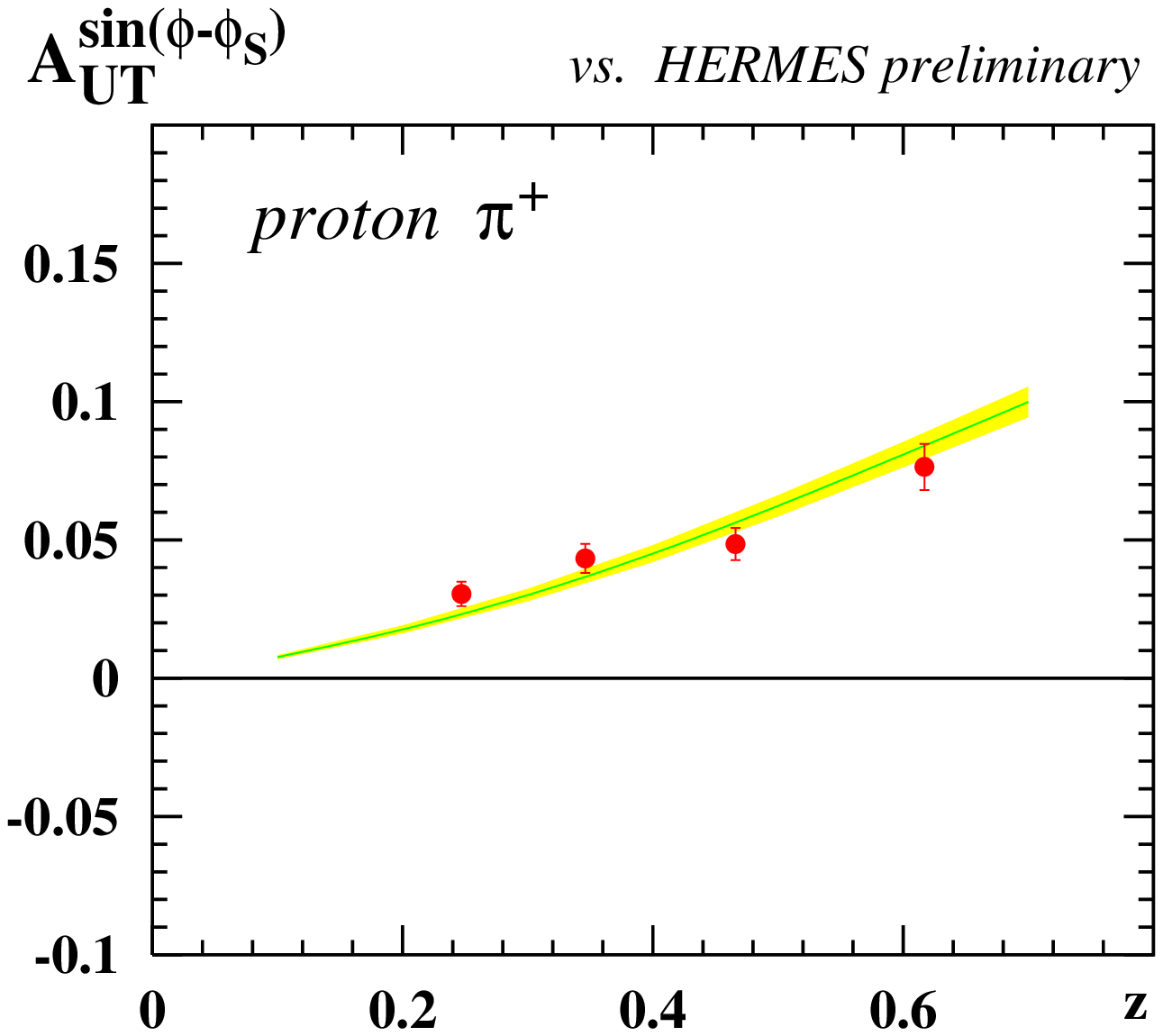} &
\includegraphics[height=3.8cm]{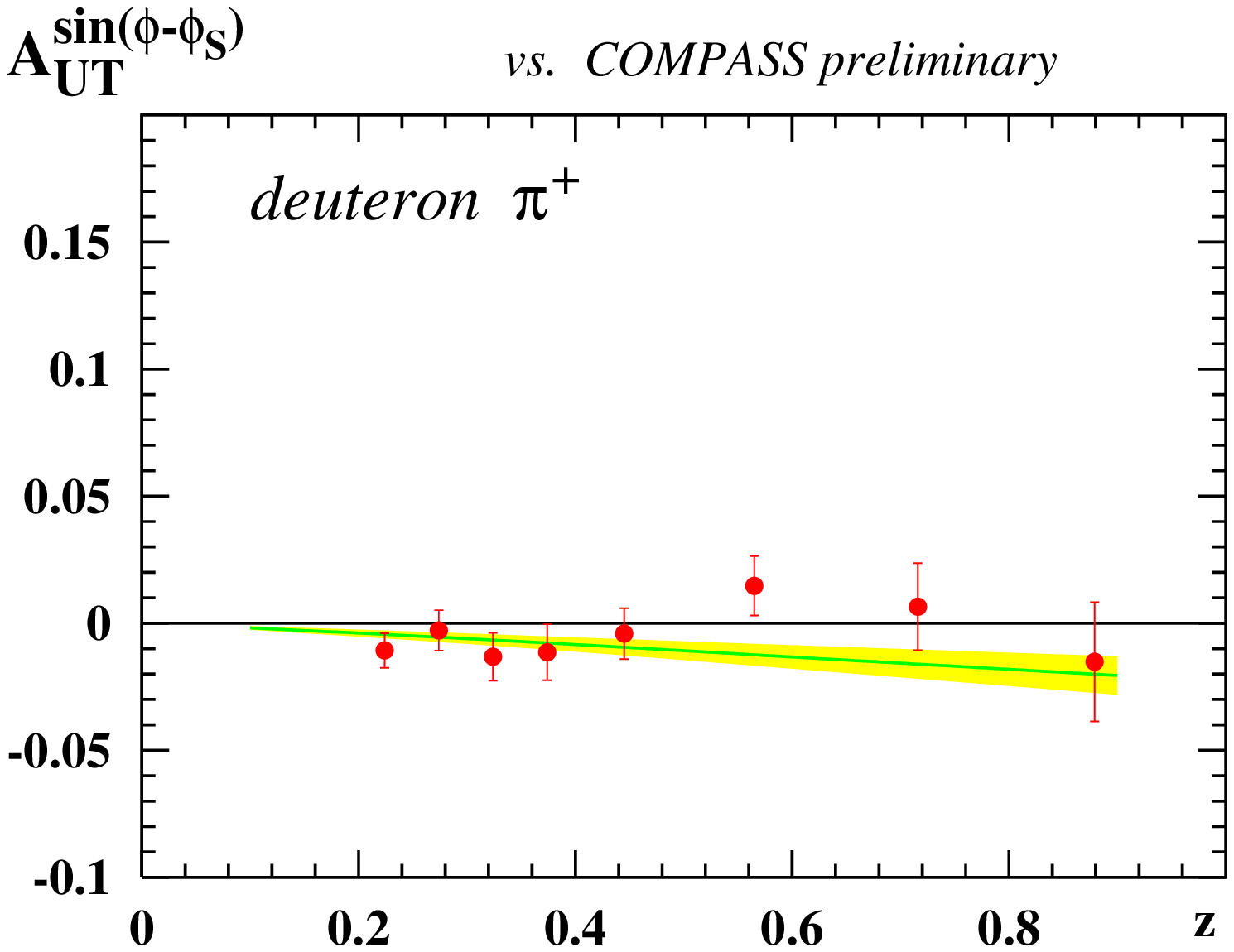}  \\
\includegraphics[height=3.8cm]{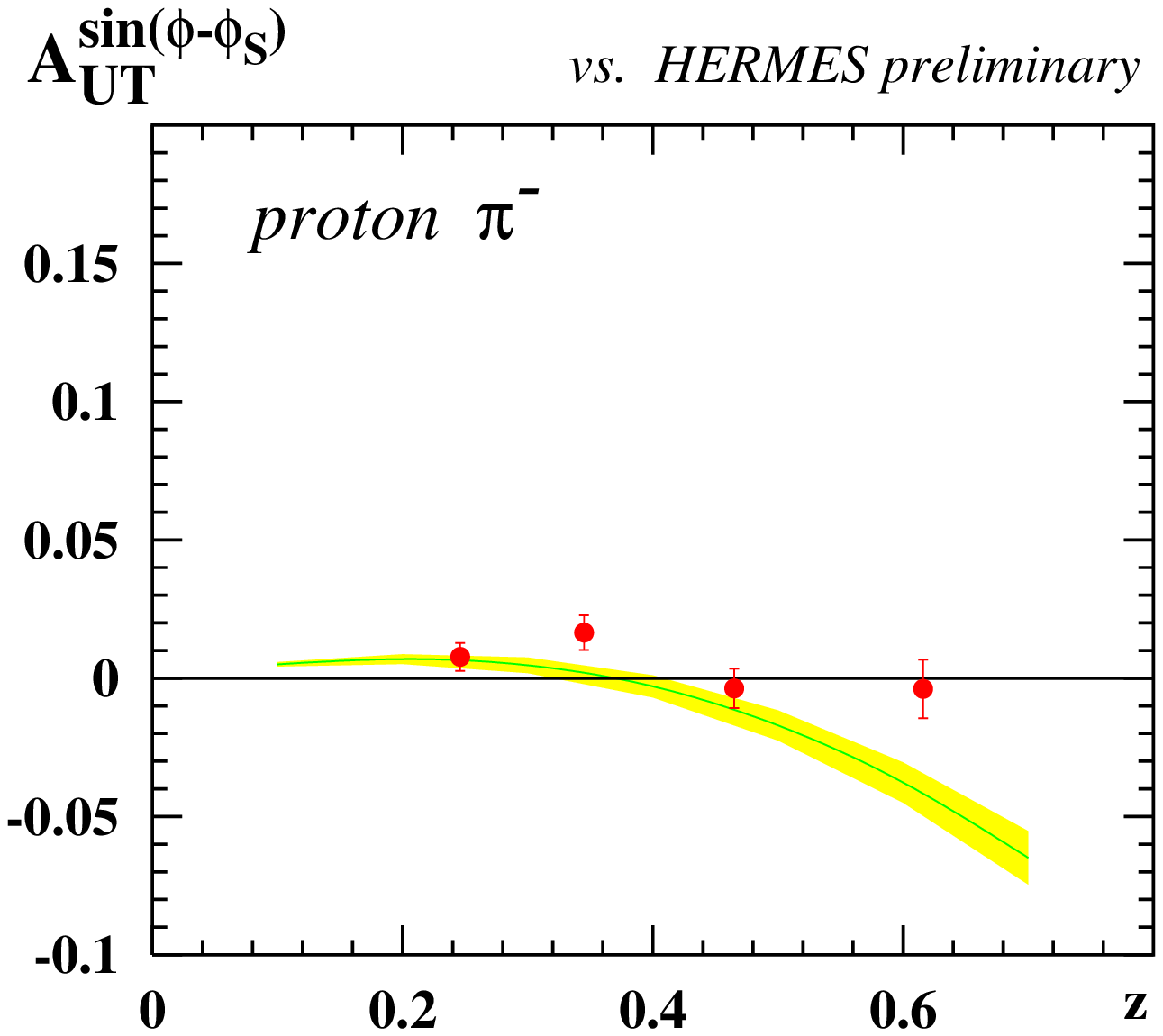} &
\includegraphics[height=3.8cm]{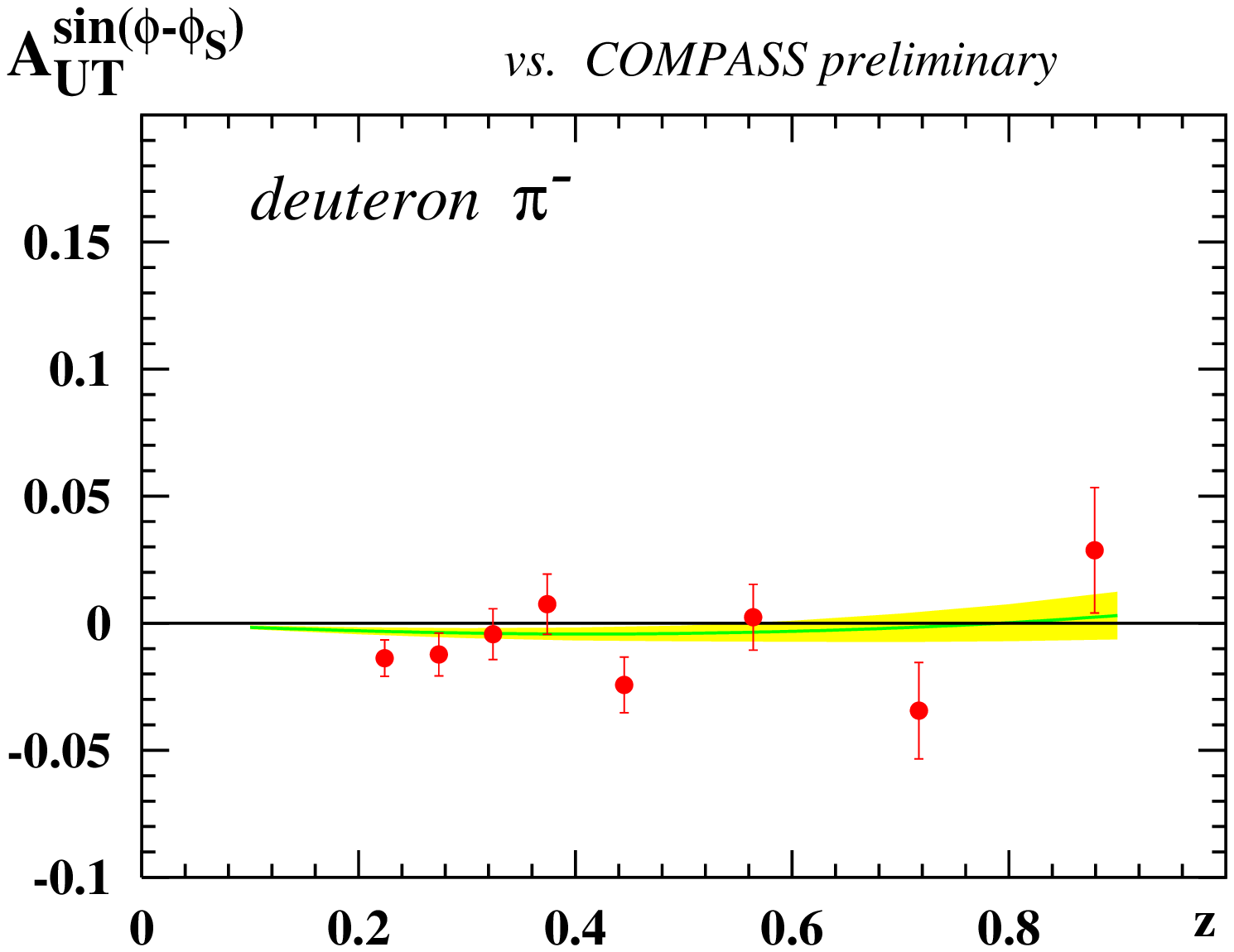} \\
\includegraphics[height=3.8cm]{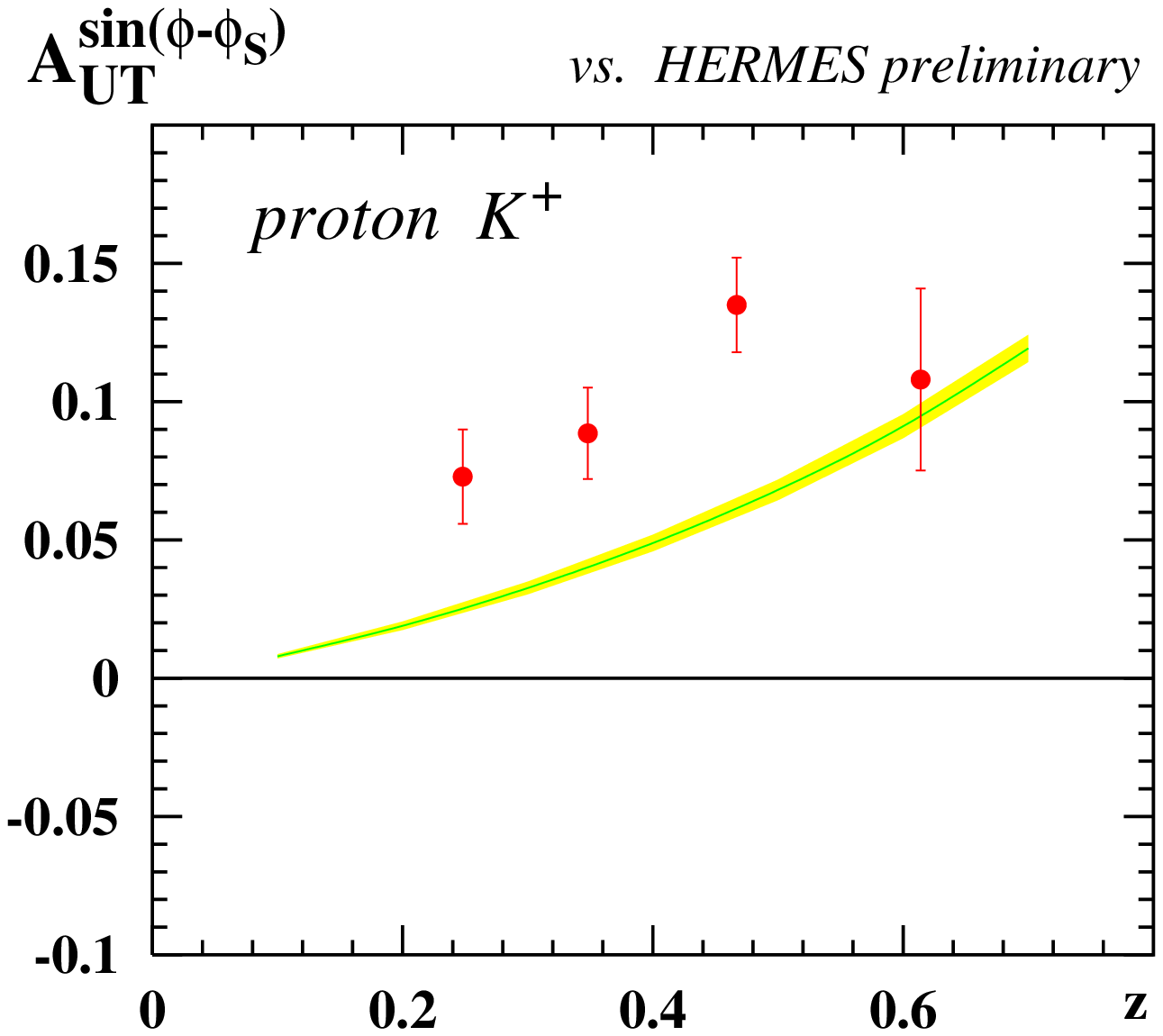}  &
\includegraphics[height=3.8cm]{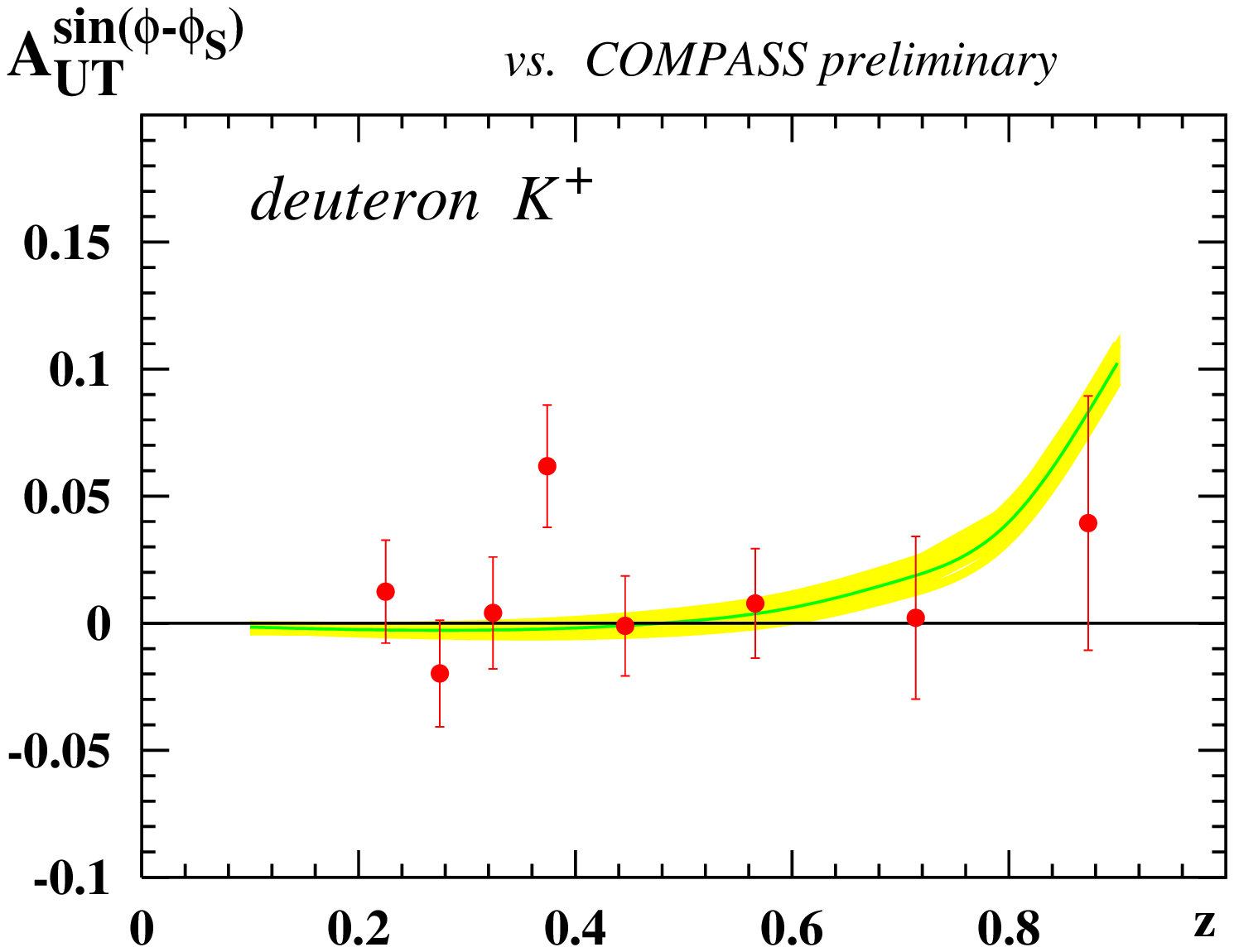} \\
\includegraphics[height=3.8cm]{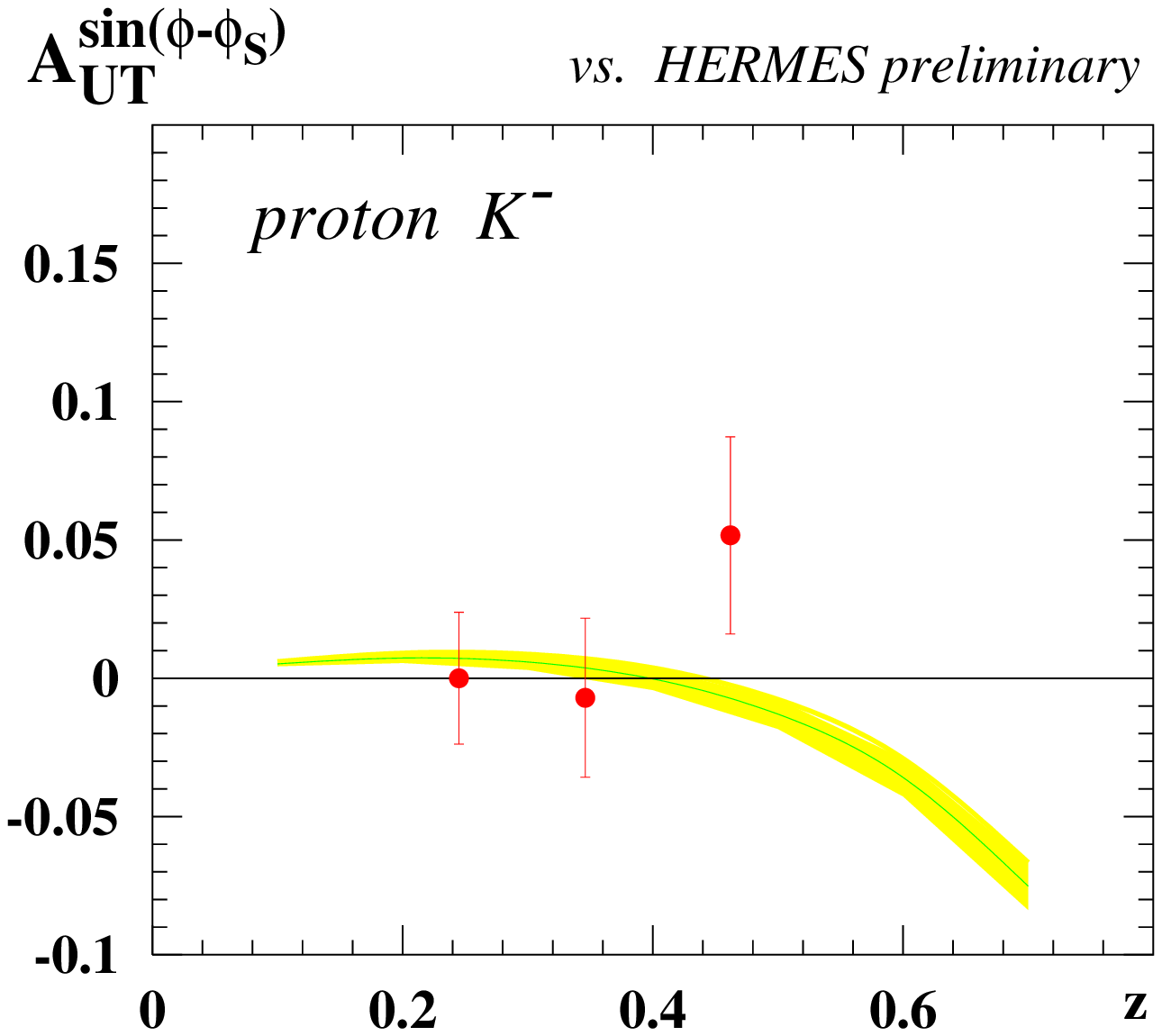}  &
\includegraphics[height=3.8cm]{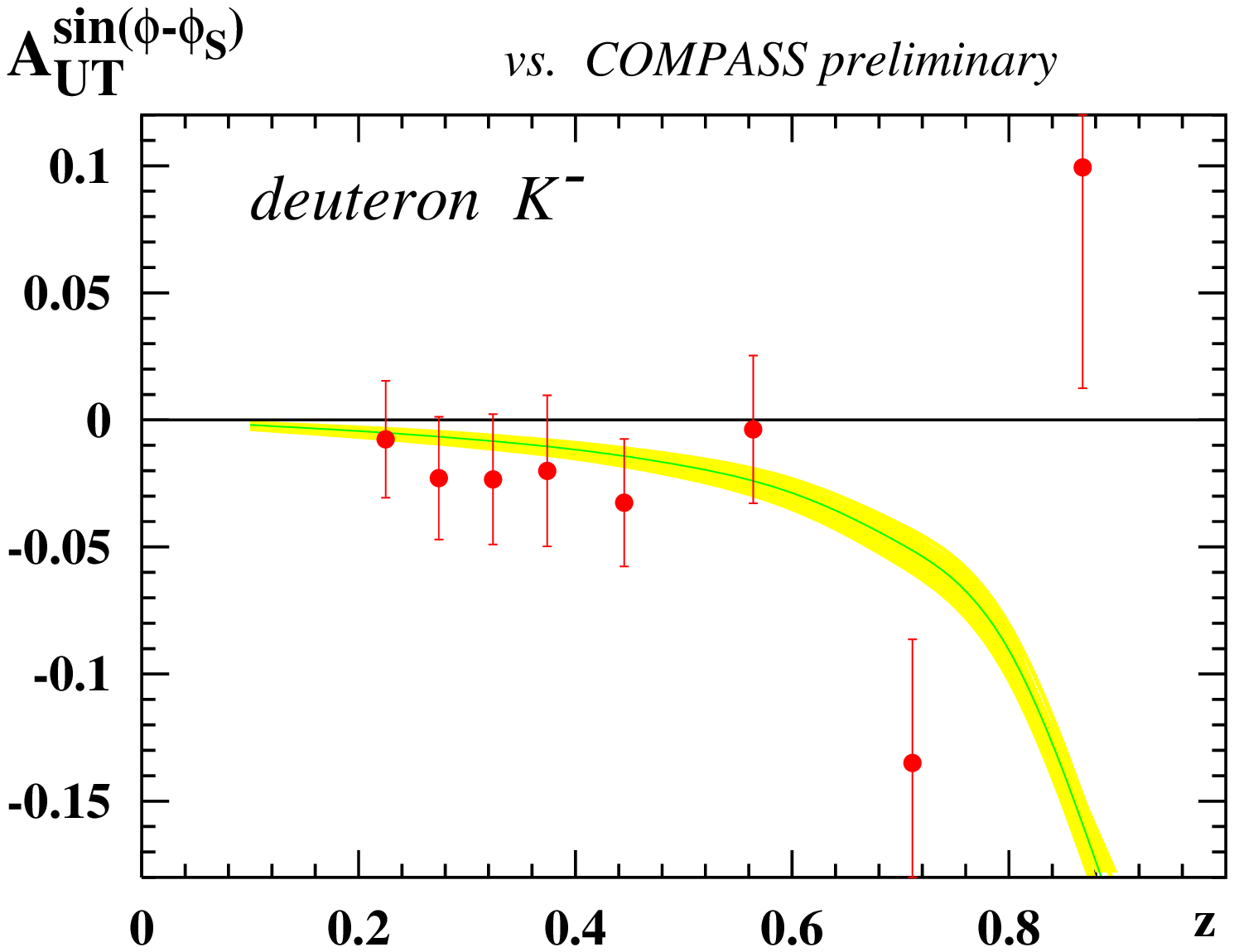} \\
\includegraphics[height=3.8cm]{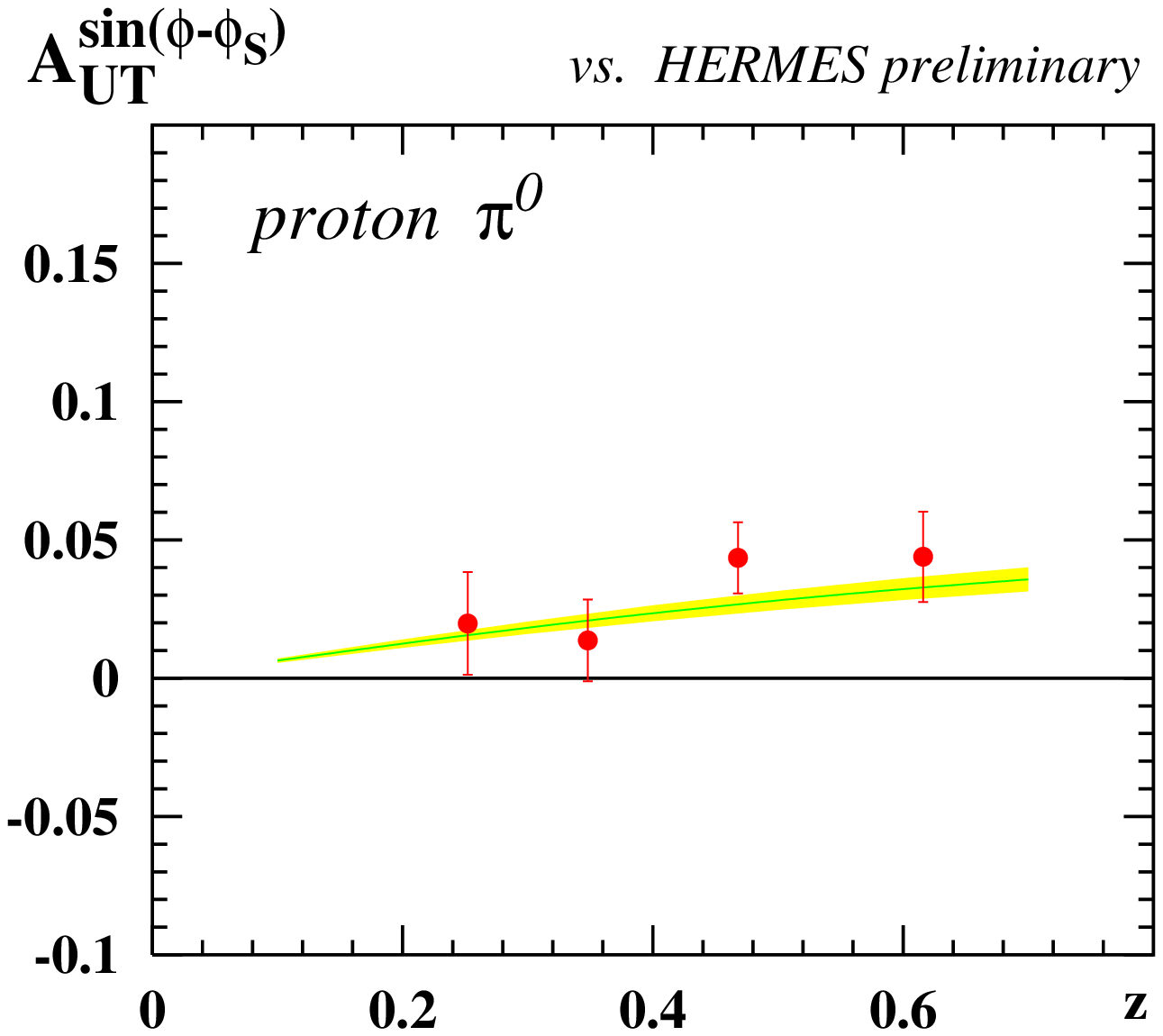} &
\end{tabular}
\caption{\label{Fig6:AUT-Siv-z}
    The Sivers SSA for various hadrons from different targets vs.\ z.
    Left  panel: HERMES data (proton target) \cite{Diefenthaler:2007rj}.
    Right panel: COMPASS data (deuteron target) \cite{Martin:2007au}.
    The theoretical curves are the best fit (solid line) and its
    1-$\sigma$-region (shaded area) as obtained from the Ansatz
    and best fit in Eqs.~(\ref{Eq:f1Tperp-ansatz},~\ref{Eq:best-fit}).
    These results are a PREDICTION of the fit
    (\ref{Eq:f1Tperp-ansatz},~\ref{Eq:best-fit}).}
\end{figure}
%

Let us postpone the discussion of this point for later, and continue
with $z$-dependence. We did not use these data in the fit. Therefore the
results, see Fig.~\ref{Fig6:AUT-Siv-z}, are a prediction of the fit
(\ref{Eq:f1Tperp-ansatz},~\ref{Eq:best-fit}), and the assumed
Gaussian model for transverse parton momenta.
The overall performance of the description is again satisfactory.
It is not surprising that the only worthwhile mentioning mismatch
is for the $K^+$ data from HERMES.

At this point, one could try to improve the description of $z$-dependence
by re-adjusting the parameters $p_{\rm Siv}^2$ and $K_{D_1}^2$ in
Eq.~(\ref{Eq:AUT-SIDIS-Gauss}). In particular,
$K_{D_1}^2$ could even be allowed to be $z$-dependent. With different
$p_{\rm Siv}^2$ and $K_{D_1}^2$ one would, of course, obtain somehow different
$A_a$ in the best fit (\ref{Eq:f1Tperp-ansatz},~\ref{Eq:best-fit}).
This could be continued to an iteration procedure, which (if convergent)
would result in a optimized description of data on $x$- and $z$-dependence.
$P_{h\perp}$-dependence could also be included.
This procedure would help to better constrain the parameters
$p_{\rm Siv}^2$ and $K_{D_1}^2$. But it presumably would have little
impact on the fit results (\ref{Eq:best-fit}).
Keeping this in mind, we shall --- at the present stage of art --- content
ourselves with the descriptions in Figs.~\ref{Fig5:AUT-Siv-x} and
\ref{Fig6:AUT-Siv-z}. The $P_{h\perp}$-dependence will be discussed elsewhere.

\section{  How does the Sivers function look like?}
\label{Sec:How-does-the-Sivers-function-look-like?}

In Sec.~\ref{Sec:Understanding-pion-and-kaon-Sivers-effect}
we have seen that the probably simplest Ansatz one can imagine
for the Sivers function, see Eq.~(\ref{Eq:f1Tperp-ansatz}), works.
'It works' means in this context that it yields an acceptable
$\chi^2$ per degree of freedom of $\chi_{\rm d.o.f.}={\cal O}(1)$.
Other, more flexible Ans\"atze could yield better descriptions.
In particular, it is an interesting question, whether one could better
describe the proton target $K^+$ Sivers effect.
Different Ans\"atze with more free parameters (and the impact of different
fragmentation functions) were explored in \cite{Melis-at-DIS-in-London},
but a 'better' description of the kaon HERMES data could not be reported
there, neither.

\begin{figure}
\begin{tabular}{ccc}
\includegraphics[height=5.8cm]{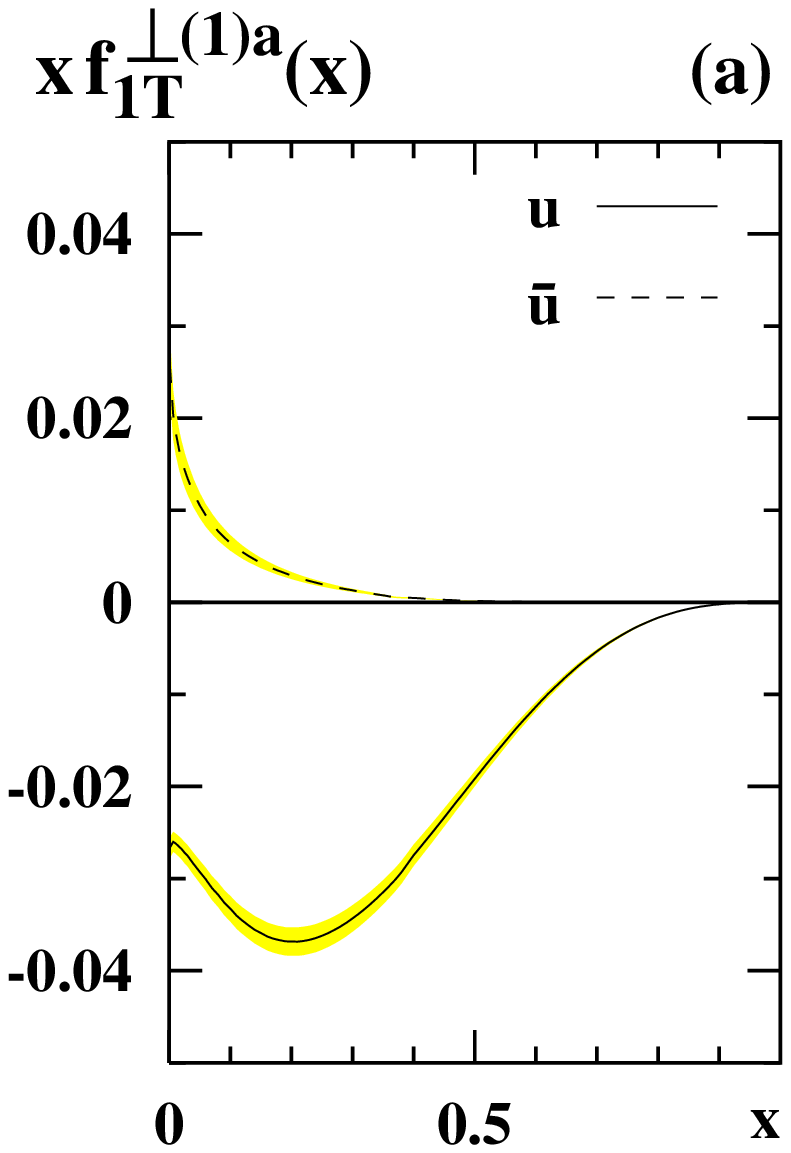} &
\includegraphics[height=5.8cm]{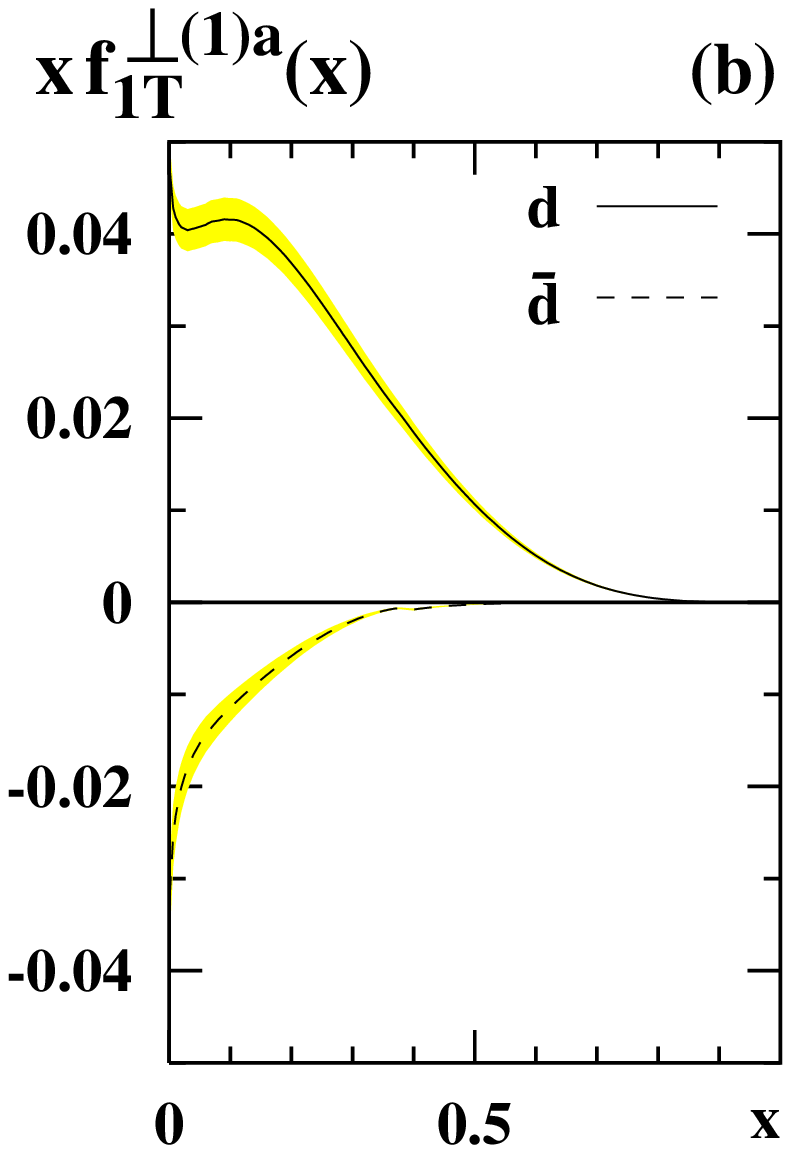} &
\includegraphics[height=5.8cm]{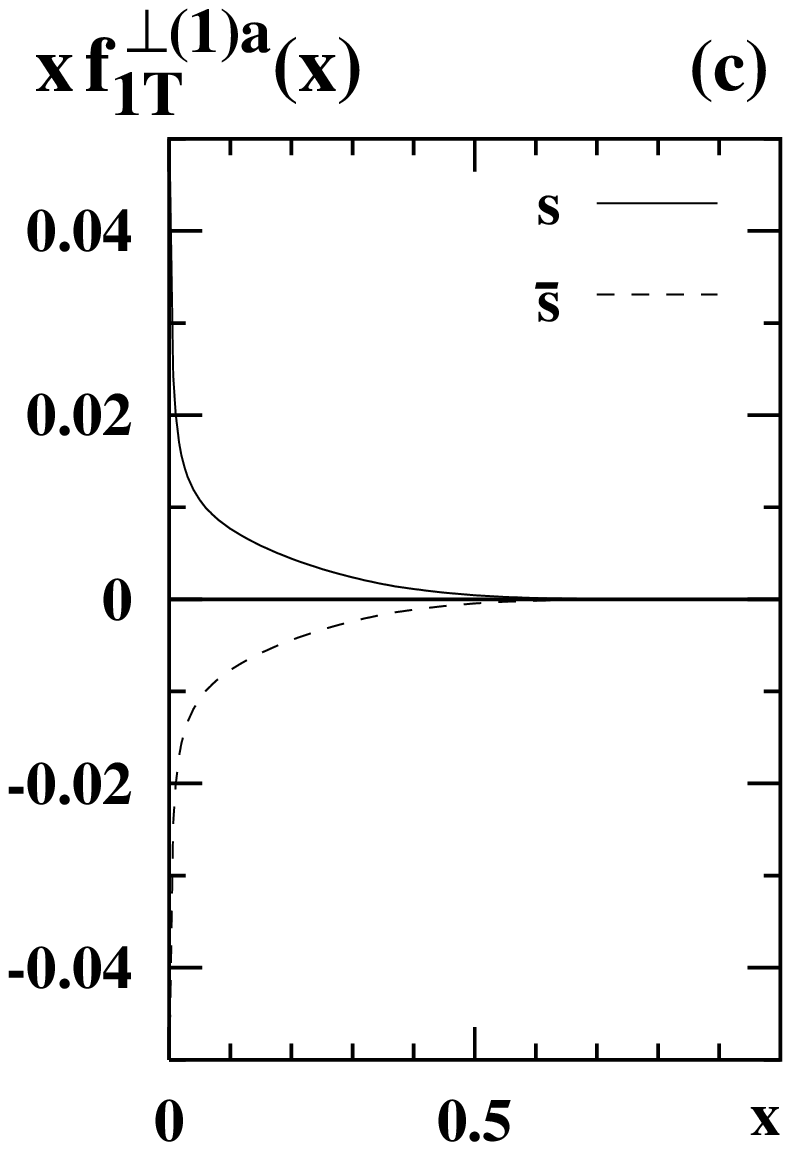}
\end{tabular}
\caption{\label{Fig7:new-Sivers-functions}
    The $xf_{1T}^{\perp(1)a}(x)$ vs.\ $x$ as
    extracted from preliminary HERMES and COMPASS
    data~\cite{Diefenthaler:2007rj,Martin:2007au}.
    (a) The flavours $u$ and $\overline{u}$.
    (b) The flavours $d$ and $\overline{d}$.
    (c) The flavours $s$ and $\overline{s}$ that were fixed
    to $\pm$ positivity bounds (\ref{Eq:Sivers-positivity}) for reasons
    explained in Sec.~\ref{Sec:Understanding-pion-and-kaon-Sivers-effect},
    see also Eqs.~(\ref{Eq:f1Tperp-ansatz},~\ref{Eq:best-fit}).
    The shaded areas in (a) and (b) show the respective
    1-$\sigma$-uncertainties.}
\end{figure}

However, having achieved a  $\chi_{\rm d.o.f.}=1.3$ one must wonder,
whether there really is a necessity to improve that fit.
Let us adopt here the point of view that it is not, and draw conclusions
from our results.
\begin{itemize}
\item   The Ansatz (\ref{Eq:f1Tperp-ansatz}) is rather rigid. It denies the Sivers
    function an independent $x$-shape, and forces it to be proportional to
    $f_1^a(x)$. The only ``freedom'' it gives to $f_{1T}^{\perp a}$ is that the
    proportionality factors $A_a$ can be flavor dependent.
    (To recall: we included the factor $\la p_T\ra_{\rm unp}/(2M_N)$ in
    (\ref{Eq:f1Tperp-ansatz}) such that $|A_a| \le 1$ guarantees
    positivity (\ref{Eq:Sivers-positivity}).)
\item   The initial, unconstrained six-parameter-fit forced the Sivers $s$
        ($\overline{s}$) function to come close to $(+1)\times$
        (to exceed $(-1)\times$) the positivity bound (\ref{Eq:Sivers-positivity}).
\item   Because of that, in the final four-parameter-fit, we fixed the
        Sivers strangeness functions such that they saturate $\pm$ the positivity
        bounds: $A_s=+1$ and $A_{\bar s} = -1$.
\item   One may worry whether such a large 'Sivers strangeness' in the nucleon
        is natural. However:
        (i) the 'net Sivers strangeness content' is zero ($s$, $\overline{s}$
        have opposite signs).
        (ii) Fixing, for example, $A_s=\frac12$ and $A_{\bar s}=-\frac12$
        (explores positivity only within~$50\%$),
        would increase the $\chi^2_{\rm d.o.f.}$ by only 0.07 units ---
        i.e.\ an equally acceptable fit. Thus, presently there is no
        reason to worry about the Sivers strangeness functions.
        (Moreover $\la p_T\ra_{\rm unp}$ in (\ref{Eq:Sivers-positivity}),
        defined within the Gauss model approximation, could be flavor-dependent
        and different for $s,\,\overline{s}$, possibly relaxing numerically
        the bound imposed here.)
\item   It is a reasonable guess that the Sivers function is suppressed at small $x$
        compared to $f_1^a(x)$ \cite{Vogelsang:2005cs}.
        But it is not necessary to build in such a suppression in the Ansatz
        in order to describe the COMPASS data rather precise at small $x$.
        In our fit this is achieved by the different relative signs of the Sivers
        functions.
\item   We do not mean that the Sivers function should raise like $f_1^a(x)$ at
        small~$x$. Rather we would like to stress that a suppression of the Sivers
        function compared to $f_1^a(x)$, if existent, is not yet constrained
        by the present data.
\item   We can draw from our study even the following stronger conclusion.
        The present data do not yet tell us much about the shape of the
        Sivers function. But they tell us something about its magnitude
        and relative signs of the different flavours.
\end{itemize}
After these cautious remarks concerning the meaning and interpretation of our
results, let us have a look how the obtained Sivers functions look like,
see Fig.~\ref{Fig7:new-Sivers-functions}.
%
%
%
%
We make the following observations.
\begin{itemize}
\item   The Sivers $u$ and $d$ distributions are of comparable magnitudes
        but opposite signs, as predicted in the large-$N_c$ limit
    \cite{Pobylitsa:2003ty}, see Eq.~(\ref{Eq:large-Nc}).
\item   The Sivers $\overline{u}$ and $\overline{d}$ distributions
    are of comparable magnitudes but opposite signs,
    as predicted in the large-$N_c$ limit \cite{Pobylitsa:2003ty},
    see text in the sequence of Eq.~(\ref{Eq:large-Nc}).
\item   It is $2A_u \approx -2A_{\bar u}\approx A_d \approx -A_{\bar d}$ and
    $A_s = -A_{\bar s}$, Eq.~(\ref{Eq:best-fit}). Were the former relations
    exact, then the contribution of $q$, $\overline{q}$ ($q=u,d,s$)
    to the Burkardt sum rule \cite{Burkardt:2003yg,Burkardt:2004ur}
        \be\label{Eq:Burkardt-sum-rule}
        \sum_{a=q,\overline{q},g}\int_0^1 dx\,f_{1T}^{\perp(1)a}(x)=0
        \ee
        would vanish, i.e.\ also $\int_0^1 dx\,f_{1T}^{\perp(1)g}(x)=0$.
        That would imply a small Sivers gluon distribution, as concluded
        independently in
        \cite{Efremov:2004tp,Brodsky:2006ha,Anselmino:2006yq}.\footnote{
    Notice that strictly speaking the integrals in (\ref{Eq:Burkardt-sum-rule})
    exist only if $A_u = -A_{\bar u}$ and $A_d = -A_{\bar d}$ holds exactly
    in the Ansatz (\ref{Eq:f1Tperp-ansatz}) (cf. in this context also the model calculation \cite{Goeke:2006ef} where quark-        and gluon Sivers functions cancel under the integral).
    For $A_u$ and $A_{\bar u}$ this is the case within 1-$\sigma$, but
    for $A_d$ and $A_{\bar d}$ there is a clearer mismatch, see
    (\ref{Eq:best-fit}), such that (\ref{Eq:Burkardt-sum-rule}) would diverge.
    However, our parameterization is not meant to be valid for $x\to0$ and
    is anyway only constraint for $x>0.003$ by COMPASS data. We remark
    that our parameterization also is not meant to be valid for $x\to1$
    where the Sivers function is expected to vanish faster than $f_1^a(x)$
    \cite{Brodsky:2006hj}.}
\item   The Sivers $u$ and $d$ distributions obtained here agree qualitatively
    with earlier works
\cite{Efremov:2004tp,Anselmino:2005ea,Vogelsang:2005cs,Collins:2005ie,Anselmino:2005an}
    in which sea quarks effects were not considered.
\item   The signs of the Sivers $u$ and $d$ distributions support the
    picture of Ref.~\cite{Burkardt:2002ks}.
\end{itemize}

\section{Predictions for CLAS AND COMPASS}
\label{Sec:predictions}

%
\begin{figure}[b!]
\begin{tabular}{cc}
\includegraphics[height=3.8cm]{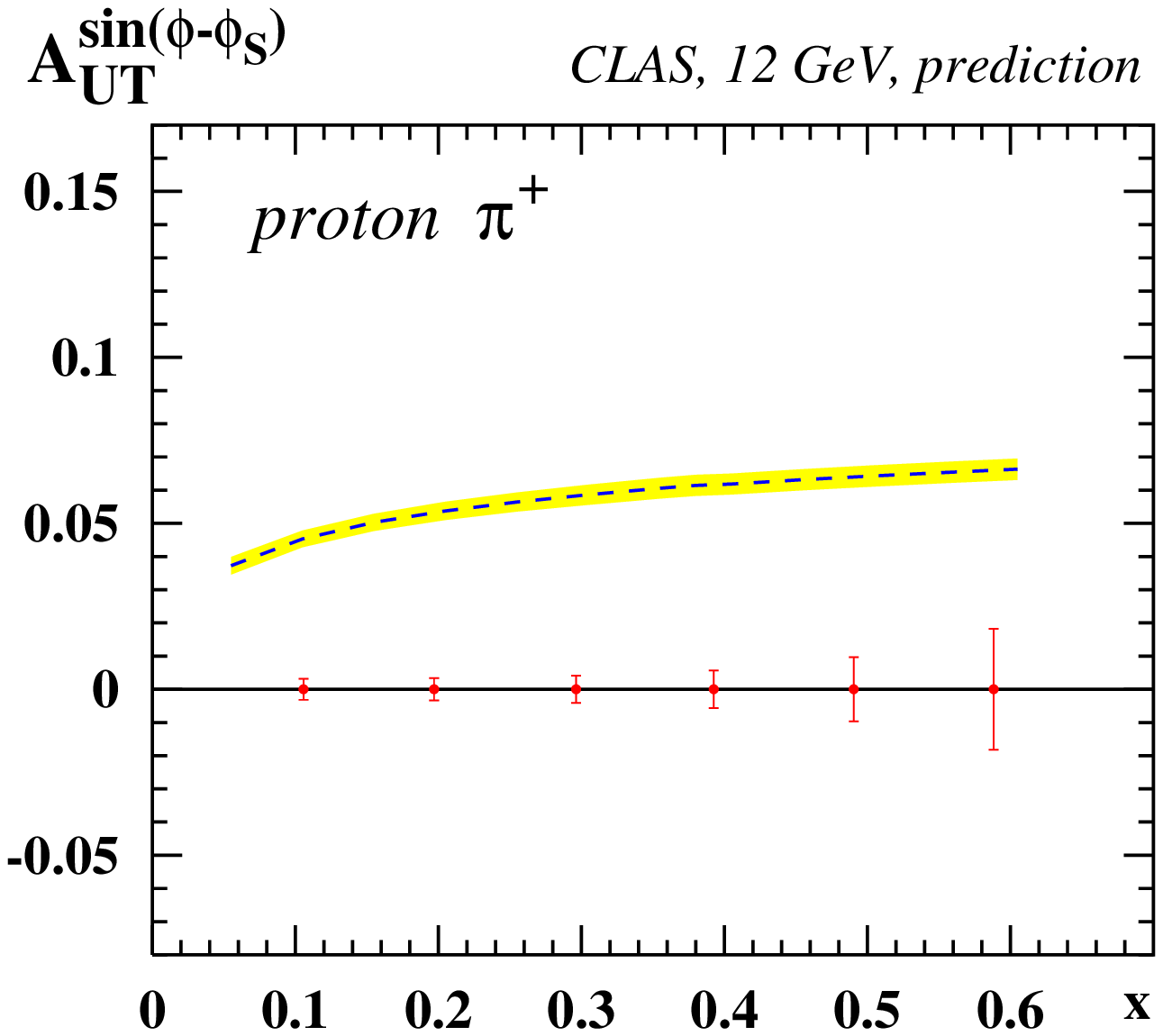} &
\includegraphics[height=3.8cm]{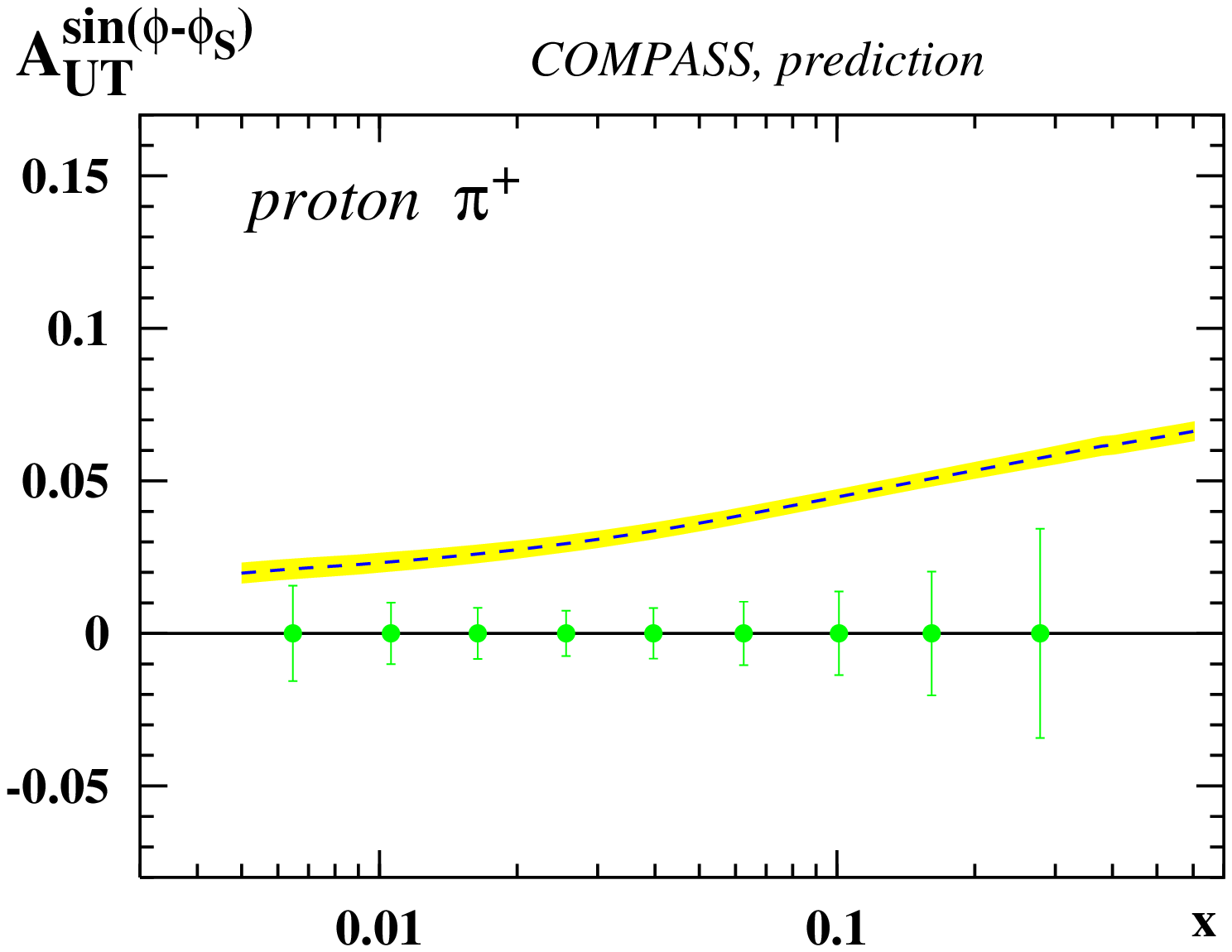}\\
\includegraphics[height=3.8cm]{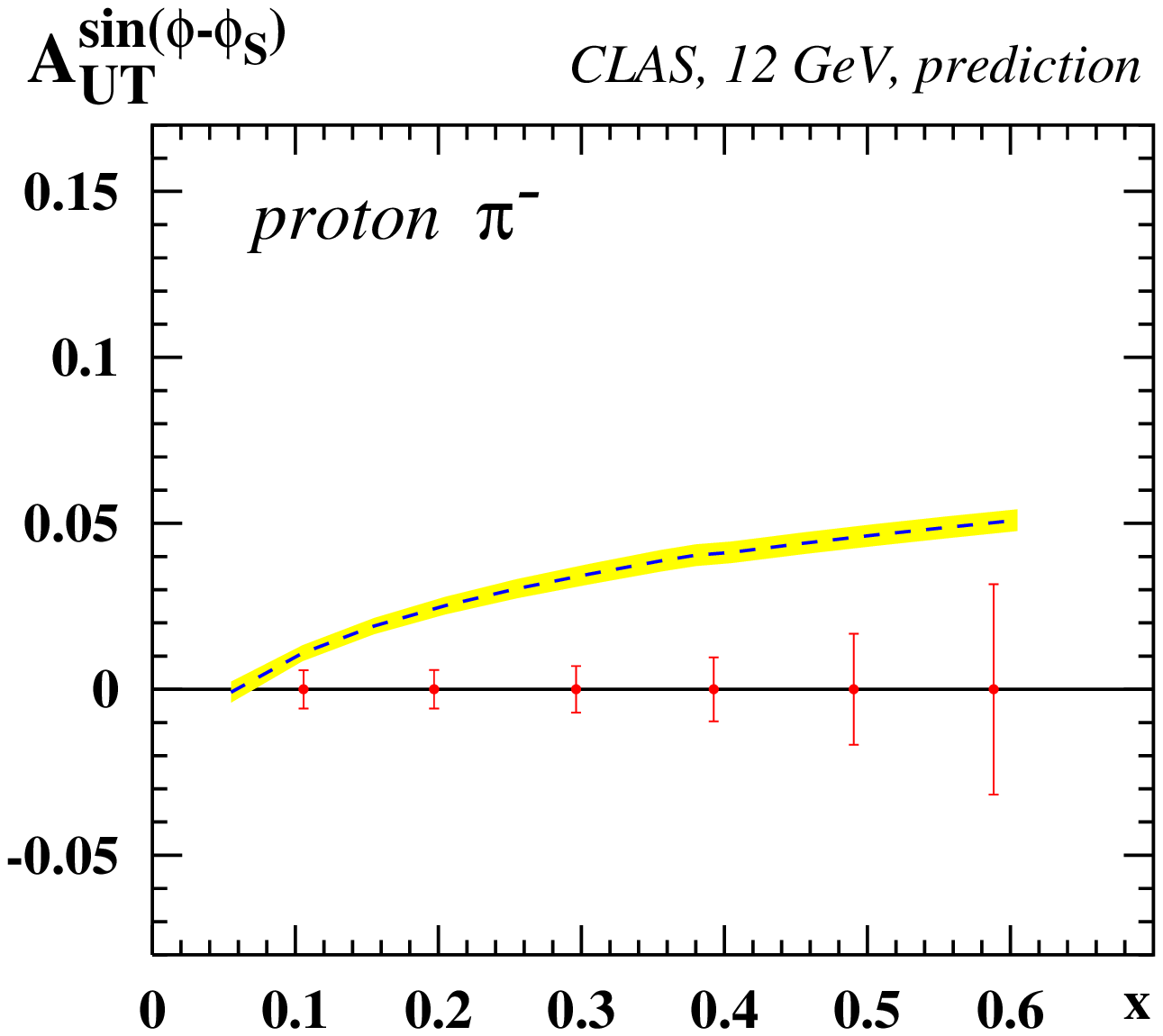} &
\includegraphics[height=3.8cm]{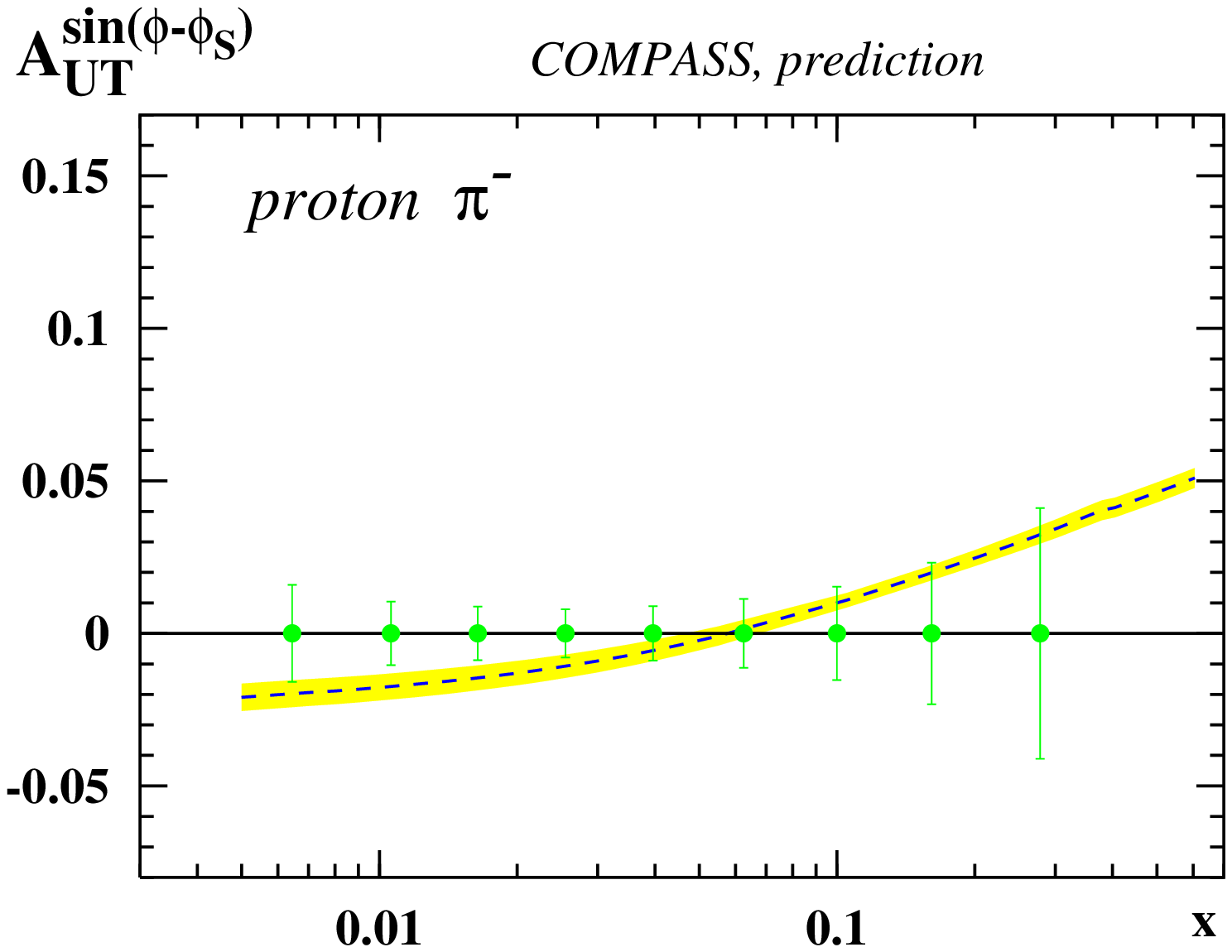}\\
\includegraphics[height=3.8cm]{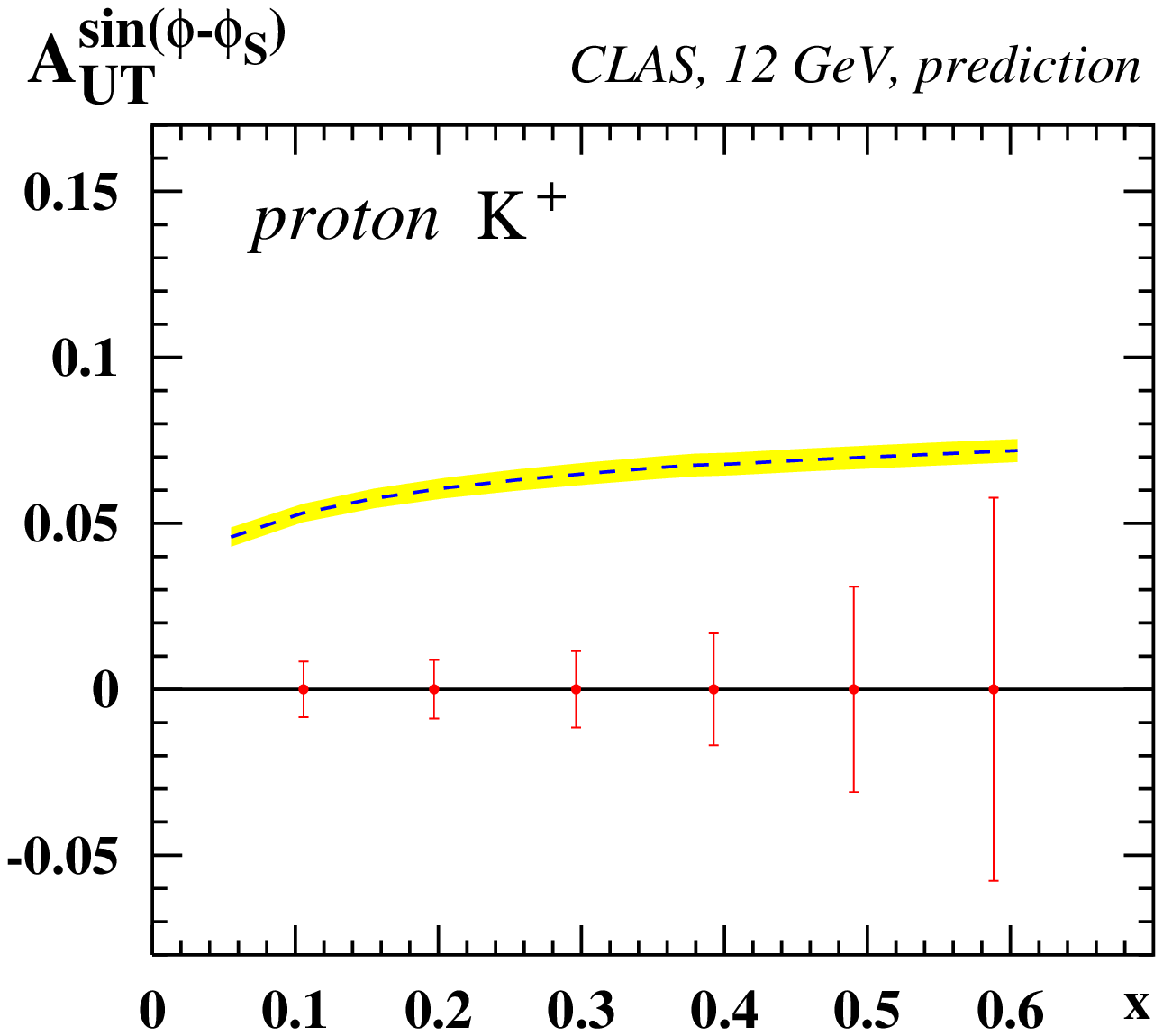} &
\includegraphics[height=3.8cm]{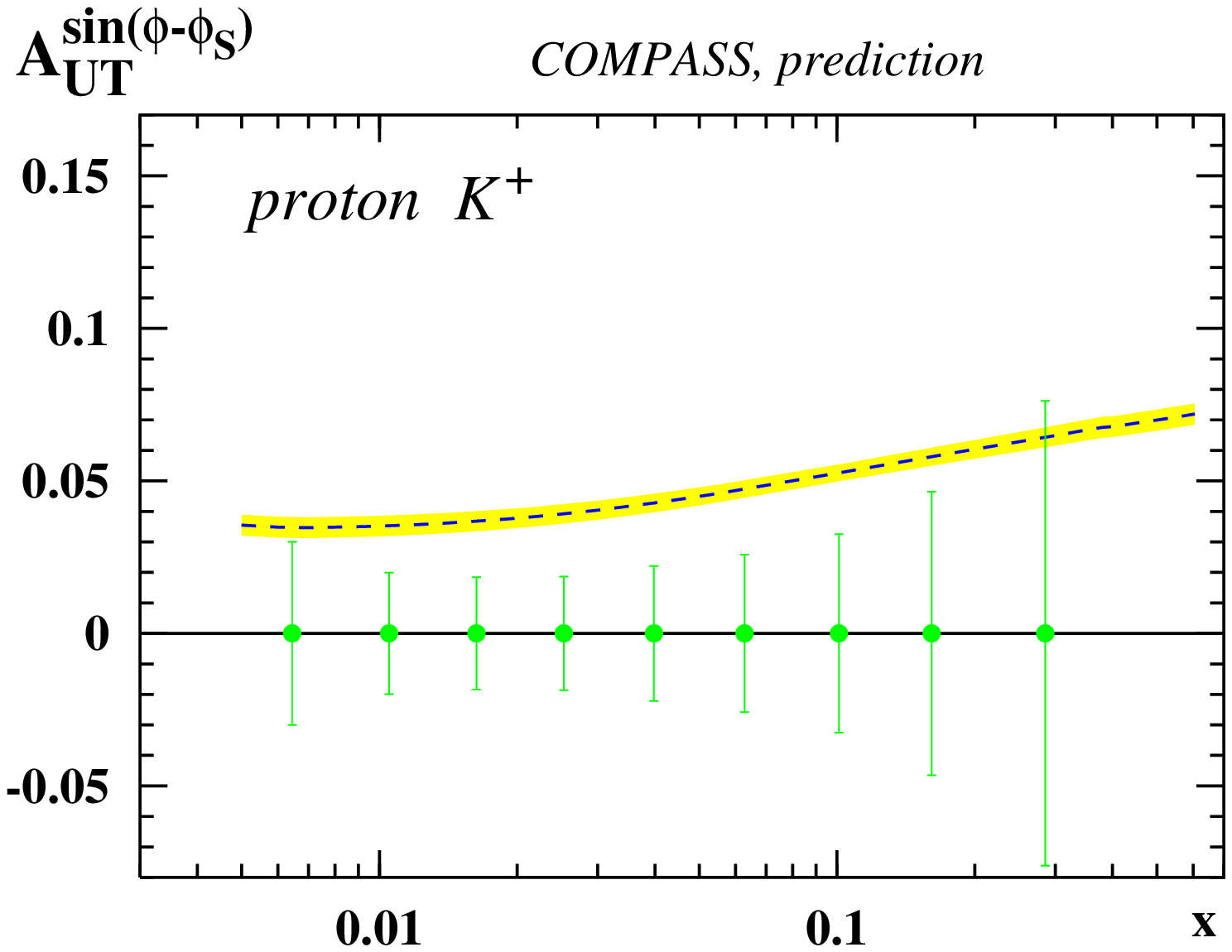}\\
\includegraphics[height=3.8cm]{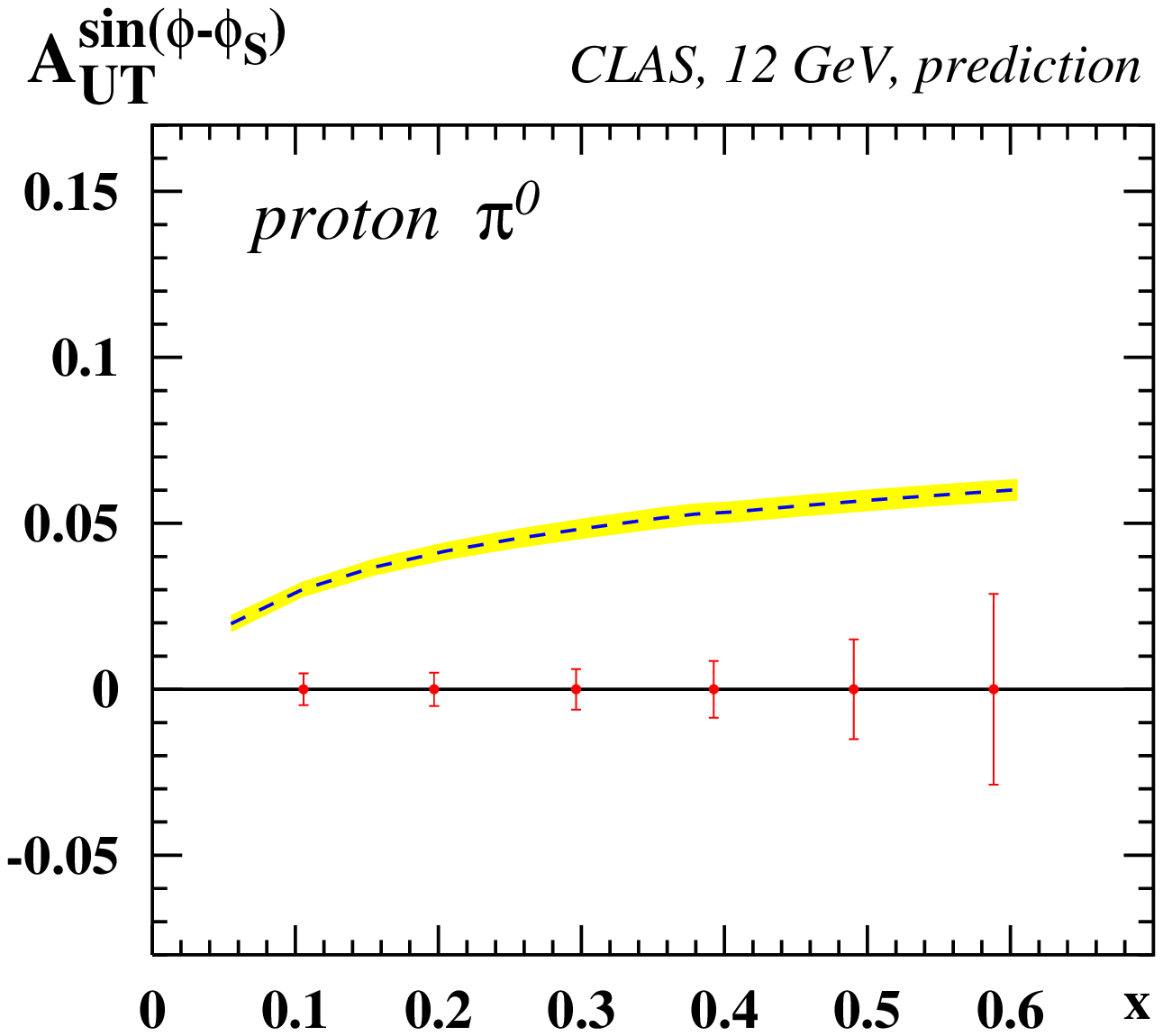} &
\includegraphics[height=3.8cm]{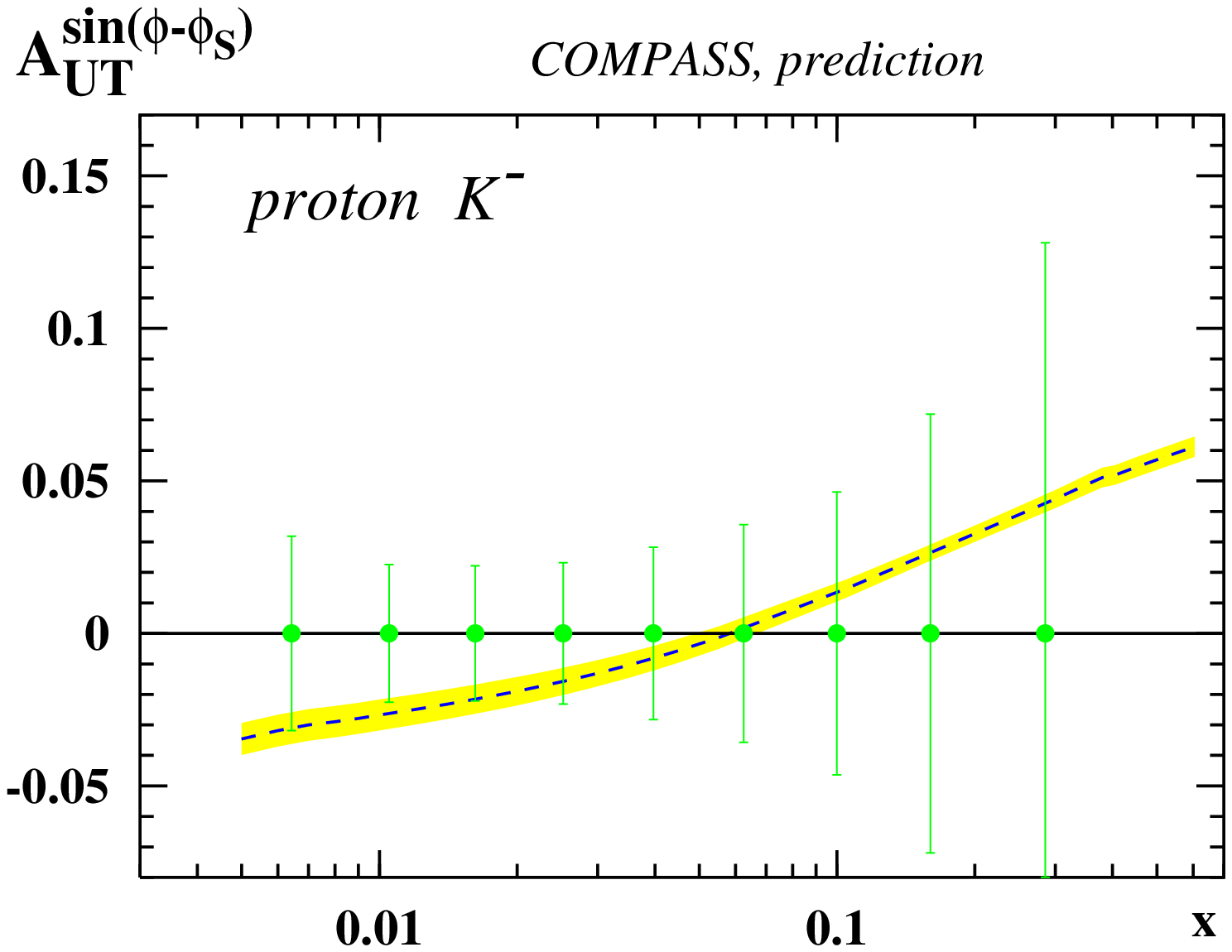}
\end{tabular}
\caption{\label{Fig8:predictions}
    Predictions for the Sivers SSA as function of $x$ from a proton target
    for CLAS at $12\,{\rm GeV}$ (left panel) and for COMPASS (right panel).
    The predictions are made on the basis of the fit
        (\ref{Eq:f1Tperp-ansatz},~\ref{Eq:best-fit}) presented in this work.
    The CLAS error projections are from 
    \cite{Harut-private-communication,JLab-LOI-CLAS12}. 
    For the COMPASS error estimates we assumed
    the statistics of \cite{Martin:2007au}.}
\end{figure}
%

Fig.~\ref{Fig8:predictions} shows the predictions, made on the basis of the fit 
(\ref{Eq:f1Tperp-ansatz},~\ref{Eq:best-fit}), for the $x$-dependence of the Sivers
SSA of charged pions and kaons from proton targets for two experiments:
CLAS with $12\,{\rm GeV}$ beam upgrade, and COMPASS. The CLAS error projections 
for 2000 hours run time are from \cite{Harut-private-communication,JLab-LOI-CLAS12}.
The COMPASS error projections are estimated assuming a statistics comparable
to that of the deuteron target experiment \cite{Martin:2007au}.

We see that both experiments will be able to confirm (or reject)
these predictions within their expected statistical accuracies.

The interesting question is, of course, whether CLAS and COMPASS will
confirm the HERMES results for $K^+$ at $x\sim 0.15$, and overshoot
our predictions in that region of $x$. In this context, however,
it is also worth to look more carefully into the kinematics.
The next Section is devoted to this task.

\section{Factorization and power corrections}

When averaged over the respectively covered kinematical regions,
CLAS, COMPASS and HERMES have a comparable $\la Q^2\ra\sim$ (2-3) ${\rm GeV}^2$
(this is what we used in our study). At fixed $x$, however,
$Q^2$ can vary significantly in these experiments. For example,
\ba
\mbox{HERMES:} & \la x\ra=0.115 \,, & \la Q^2\ra=2.62\,{\rm GeV}^2, \nonumber \\
\mbox{COMPASS:}& \la x\ra=0.1205\,, & \la Q^2\ra=12.9\,{\rm GeV}^2,
    \label{Eq:compare-kinematics}
\ea
which is in the $x$-region where the 'trouble' occurs.
(At CLAS it is about $2\,{\rm GeV}^2$.)

Being sure that one really deals with the leading twist contribution, such differences
are not dramatic --- as long as we are not interested in a high precision study of the
effect. But how can we {\sl a priori} be sure that there are no power corrections?
Little is known about such corrections to the Sivers effect. For illustrative
purposes, let us assume that
\be\label{Eq:power-corrections}
    A_{UT\,\rm measured}^{\sin(\phi-\phi_S)}
    = \biggl\{
      \mbox{'twist-2 Sivers effect' in Eqs.~(\ref{Sivers},~\ref{Eq:AUT-SIDIS-Gauss})}
      \biggr\}
    + C(Q)\,\frac{M_N^2}{Q^2} \;\hspace{0.5cm}
\ee
as one generically may expect.
The 'coefficient' $C(Q)$ could, in general, be flavour-dependent and typically
depend on scale logarithmically (and depend on $x$, $z$, $\dots$ etc.).
By looking at (\ref{Eq:compare-kinematics},~\ref{Eq:power-corrections})  we see,
that power corrections --- if they play a role --- are about 5 times smaller
at COMPASS compared to HERMES at $x\sim 0.15$.

Maybe such corrections are irrelevant for $Q^2 > 1\,{\rm GeV}^2$
which is typically used as DIS-cut.
In any case, a careful comparison of all (present and future) data
from COMPASS, HERMES and JLab will shed light on the possible size
of power corrections.

In particular, data from CLAS (first transverse target data will be available in 2011 
\cite{JLab-proposal-CLAS6}) will be valuable --- were a wide kinematical range
is covered with very high statistics. Moreover, even after the $12\,{\rm GeV}$
beam energy upgrade, a certain fraction of the beam time also somehow
lower beam energies will be available. Therefore, CLAS could provide
precise information on SSAs at fixed value of $x$ but for different $Q^2$.
Such data would allow to discriminate between 'leading twist' contributions
and possible power corrections --- paving the way to a concise understanding
of the novel effects.

One should not forget that in general 'power corrections'
are by no means only 'contaminations' to the parton model formalism.
Rather, they also contain interesting physics, we refer to \cite{Chen:2005tda}
for an example.

\section{Conclusions}

Since the release of first data on Sivers effect in SIDIS  by HERMES
and COMPASS \cite{Airapetian:2004tw,Alexakhin:2005iw}, more
detailed, higher statistics (published and preliminary) data became available
\cite{Diefenthaler:2005gx,Ageev:2006da,Diefenthaler:2006vn,Diefenthaler:2007rj,Martin:2007au,Vossen:2007mh,Alekseev:2008dn}.
While the first data did not allow to constrain Sivers sea quarks
\cite{Efremov:2004tp,Anselmino:2005ea,Vogelsang:2005cs,Collins:2005ie,Anselmino:2005an},
it is clear that the most recent ones \cite{Diefenthaler:2007rj,Martin:2007au}
can only be described, if the Sivers sea quarks are not zero.

In this work we studied the recent preliminary data \cite{Diefenthaler:2007rj,Martin:2007au}.
For that we introduced relevant Sivers sea quarks
($\overline{u}$, $\overline{d}$, $s$, $\overline{s}$),
and choose a simple Ansatz that assumes the Sivers functions to be proportional to the
unpolarized ones. The statistically satisfactory fit provides a good {\sl overall}
description of the preliminary data \cite{Diefenthaler:2007rj,Martin:2007au}.

The resulting Sivers $u$, $d$, $\overline{u}$, $\overline{d}$ distributions are
in excellent agreement with large $N_c$ predictions \cite{Pobylitsa:2003ty}.
The Sivers $u$, $d$ distributions confirm earlier analyses
\cite{Efremov:2004tp,Anselmino:2005ea,Vogelsang:2005cs,Collins:2005ie,Anselmino:2005an},
and support the physical picture of the Sivers effect from \cite{Burkardt:2002ks}.
The fit favors Sivers $s$, $\overline{s}$ distributions of opposite sign,
and close to positivity bounds (in the model-dependent way we implemented them),
though this observation has less statistical significance.

That such a fit, that assumes $f_{1T}^{\perp(1)a}(x)\propto f_1^a(x)$ for all flavours,
works implies the following. The present data do not yet give much insights into 
details of the shapes of the Sivers functions. But they already tell us something 
about their magnitudes and relative signs, and this situation will improve due to 
the impact of future data.

One could be happy about this situation. Earlier studies and the optimism, concerning
measurability of the Sivers effect in DY, are confirmed
\cite{Efremov:2004tp,Anselmino:2005ea,Vogelsang:2005cs,Collins:2005ie,Anselmino:2005an}. 
The results fit into theoretical expectations \cite{Pobylitsa:2003ty,Burkardt:2002ks}. 
There is a rough, qualitative agreement with model results
\cite{Yuan:2003wk,Bacchetta:2003rz,Lu:2004au,Cherednikov:2006zn,Gamberg:2007wm,Courtoy:2008vi}
especially concerning Sivers quark distributions.
But there is also a grain of salt.

The description of the preliminary data \cite{Diefenthaler:2007rj} on the $K^+$
Sivers effect around $x\sim 0.15$ is not ideal. Actually, we speak about 2 out 
of 61 data points\footnote{
    We mean here data on $x$-dependence from \cite{Diefenthaler:2007rj,Martin:2007au}
    (data on neutral kaons from \cite{Alekseev:2008dn} are not included in this 
    counting).
    The relevant events, when binned correspondingly, also 'pop up' in data on,
    for example, the $z$-dependence and cause 'mismatches' also there, see
    Sec.~\ref{Sec:Understanding-pion-and-kaon-Sivers-effect}.
    But these are not 'additional mismatches', since the different data sets are 
    correlated.}
available presently on the Sivers effect in DIS production of various hadrons from 
various targets. Could these two data points be statistical fluctuations? Should 
they make us worry? Could they indicate power corrections? Or do they hint at 
novel effects?

At the present stage it is not possible to draw definite conclusions.
Unfortunately, the HERMES experiment is terminated.
But fortunately there are promising prospects due to COMPASS with a proton target, 
and JLab where different transversely and longitudinally polarized targets will be 
explored \cite{JLab-proposal-CLAS6}.

The data from these experiments, which are complementary from the point of view of
kinematics, will help to answer these and many other questions --- on the Sivers
and many other effects
\cite{Vogelsang:2005cs,Efremov:2006qm,Anselmino:2007fs,Kotzinian:2006dw,Avakian:2007mv,Gamberg:2007wm,Barone:2008tn}.

\newpage


\begin{theacknowledgments}
We thank the organizers of the ``CLAS 12 RICH Detector Workshop''
(see {\tt http://conferences.jlab.org/CLAS12/index.html}) for giving us~the 
opportunity to present and discuss our recent work.\\
\\
We thank Harut Avakian for providing the error projections shown
in Fig.~\ref{Fig8:predictions} and valuable comments on the manuscript.
We also thank
John Collins, Kyungseon Joo, Andreas Metz, 
Pavel Pobylitsa and Oleg Teryaev for useful discussions.
This work is  supported by BMBF (Verbundforschung), COSY-J\"ulich
project, the Transregio Bonn-Bochum-Giessen, and is part of the by
EIIIHT project under contract number RII3-CT-2004-506078. A.E.\ is
also supported by RFBR grant 06-02-16215, by RF MSE
RNP.2.2.2.2.6546 (MIREA) and by the Heisenberg-Landau Program of
JINR.\\
\\
Notice: Authored by Jefferson Science Associates, LLC under U.S. DOE Contract No. DE-AC05-06OR23177. The U.S. Government retains a non-exclusive, paid-up, irrevocable, world-wide license to publish or reproduce this manuscript for U.S. Government purposes.

\end{theacknowledgments}

\bibliographystyle{aipproc}   

\bibliography{ProcKaonSivers-02d}

\end{document}